\documentclass[draft,aps,twocolumn,showpacs,floats,
superscriptaddress]{revtex4}  
\usepackage{amsmath}
\usepackage{amssymb}
\usepackage{graphicx, psfig}
\newcommand{\lsim}{\buildrel<\over{_\sim}}
\newcommand{\Br}{{\rm Br}}
\newcommand{\mtbfit}{M_{tb}^{\rm fit}}
\newcommand{\mtbw}{M_{tb}^{\rm w}}
\newcommand{\mtb}{m_{tb}}

\newcommand{\stp}{{\tilde{t}_1}}
\newcommand{\mstp}{{m_{\tilde{t}_1}}}

\newcommand{\sbt}{{\tilde{b}_1}}
\newcommand{\sbti}{{\tilde{b}_i}}
\newcommand{\msbt}{m_{\tilde{b}_1}}

\newcommand{\msbts}{m_{\tilde{b}_2}}
\newcommand{\glu}{{\tilde{g}}}
\newcommand{\mglu}{m_{\tilde{g}}}
\newcommand{\chapm}{{\tilde{\chi}^{\pm}_1}}

\newcommand{\chap}{{\tilde{\chi}^+_1}}
\newcommand{\chaspm}{{\tilde{\chi}^{\pm}_2}}

\newcommand{\neu}{{\tilde{\chi}^0_1}}
\newcommand{\neus}{{\tilde{\chi}^0_2}}
\newcommand{\neuth}{{\tilde{\chi}^0_3}}
\newcommand{\neufr}{{\tilde{\chi}^0_4}}

\newcommand{\nedge}{N_{\rm edge}}
\newcommand{\Bredge}{{\rm Br(edge)}}
\newcommand{\nall}{N_{\rm all}}
\newcommand{\nfit}{N_{\rm fit}}
\newcommand{\nprod}{N_{\rm prod}}
\begin{document}

\preprint{ICRR-Report-497-2003-1}

\preprint{YITP-03-17}
\title{A Detailed Study of the Gluino 
Decay into the Third Generation Squarks \\at the CERN LHC
}
\author{Junji Hisano} 
\affiliation{ICRR, University of Tokyo,
Kashiwa 277-8582, Japan }
\author{Kiyotomo Kawagoe}
\affiliation{Department of Physics,
Kobe University, Kobe 657-8501, Japan}
\author{Mihoko M. Nojiri}
\affiliation{YITP, Kyoto University, Kyoto 606-8502, Japan}
\date{\today}

\begin{abstract}
In supersymmetric models  
a gluino can decay into $tb\tilde{\chi}^{\pm}_1$  
through a stop or a sbottom.  The decay chain  
produces an edge structure in the $m_{tb}$ distribution.
Monte Carlo simulation studies show 
that the end point and the 
edge height would be measured at the CERN LHC by using a 
sideband subtraction technique. 
The stop and sbottom masses
as well as their decay  branching ratios  
are constrained by the measurement. 
We study interpretations
of the measurement in the minimal supergravity model.  
We also study
the gluino decay into $tb\tilde{\chi}^{\pm}_2$  
as well as the influence of the stop left-right mixing
on the $m_{bb}$ distribution of the tagged $tb$ events.
\end{abstract}

\pacs{\bf 12.60.Jv, 14.80.Ly 
} 

\maketitle

\section{Introduction}
The minimal supersymmetric standard model (MSSM) is one of the
promising extensions of the standard model  \cite{SUGRA}. 
The model requires
superpartners of the standard model  particles (sparticles), 
and the large hadron collider (LHC) at CERN
might confirm the existence of the new particles \cite{TDR}. 
The LHC is a $pp$ collider at $\sqrt{s}=14$ TeV, whose operation is
currently expected to start in 2007.
The integrated luminosity will be 10~fb$^{-1}$/year at the beginning
(low luminosity runs),
and then upgraded to 100~fb$^{-1}$/year (high luminosity runs). 

Supersymmetry (SUSY) must be broken and the sparticle mass spectrum depends
on the SUSY breaking mechanism.
Measurement of the sparticle masses provides a way to probe
the origin of the SUSY
breaking in nature. The sparticle mass measurement at the LHC 
has been therefore 
extensively studied. 

Among the sparticles
the third generation squarks,
stops ($\tilde{t}_i$) and sbottoms ($\tilde{b}_i$) ($i=1,2$),
get special imprints 
from physics at the very high energy scale. 
One of the examples is found in a 
model with universal scaler masses.  Although the scalar 
masses are universal at a high energy scale, the third generation 
squarks are much lighter than the first and second generation 
squarks due to the Yukawa running effect. On the other hand, 
some SUSY breaking models, such as the flavor U(2)
model \cite{u2model} or the decoupling solution \cite{decoupling}, and
the superconformal model \cite{Nelson:2000sn} for the SUSY flavor problem,
have 
non-universal boundary conditions for 
the third generation mass parameters
at the GUT scale. 
In addition, the $\tilde{t}_L$-$\tilde{t}_R$-Higgs trilinear coupling
$A_t$ is comparable to the gluino mass at the weak scale when the
origin of the SUSY breaking comes from the GUT scale or the Planck
scale physics.  This appears in the stop mass matrix with a large
coefficient $m_t$; $m_{LR}^2= m_t(A_t-\mu\cot\beta)$. The stop
left-right mixing is therefore expected to be sizable, leading to an
even lighter stop mass compared to those of other squarks \cite{infrared}.
It should be stressed that the stop
masses and the mixing are very important parameters to predict the
light Higgs mass
\cite{higgsmass}, or the rare $B$ decay ratios \cite{Barbieri:1993av}.

We may be able to access the nature of the stop and sbottom at the LHC
provided that 
they are lighter than the gluino ($\tilde{g}$). 
They may copiously arise from the gluino decay if 
the decay is kinematically allowed.
The relevant decay modes for $\tilde{b}_i$ ($i=1,2$), $\tilde{t}_1$,
to charginos $\tilde{\chi}_j^{\pm}$ ($j=1,2$) or neutralinos 
$\tilde{\chi}^0_j$ ($j=1,2,3,4$) are  
listed below (indices to distinguish a particle and its anti-particle 
is suppressed unless otherwise stated)\footnote{We do not consider a gluino decay into $t\tilde{t}_2$
for this paper.} ,
\begin{eqnarray}
{\rm (I)}_j&~&\glu\rightarrow b\sbt
\rightarrow bb\tilde{\chi}^0_j
\ (\rightarrow bbl^+l^-\tilde{\chi}^0_1),\cr
{\rm (II)}_j&~&\glu\rightarrow t\stp \rightarrow tt\tilde{\chi}^0_j,\cr
{\rm (III)}_j&~&\glu\rightarrow t\stp \rightarrow 
tb\tilde{\chi}^{\pm}_j,\cr
{\rm (III)}_{ij}&~&\glu\rightarrow
b \tilde{b}_i \rightarrow  b W\stp \rightarrow 
bbW\tilde{\chi}^{\pm}_j,\cr
{\rm (IV)}_{ij}&~&\glu\rightarrow b \tilde{b}_i\rightarrow tb 
\tilde{\chi}^{\pm}_j.
\label{gluinodecay}
\end{eqnarray}
In previous literatures~\cite{TDR,Hinchliffe:1996iu}, the lighter 
sbottom $\tilde{b}_1$
is often studied through the mode (I)$_2$, namely the
$bb\tilde{\chi}^0_2\rightarrow bbl^+l^-\tilde{\chi}^0_1$
channel. This mode is important when the second lightest neutralino 
$\tilde{\chi}^0_2$ has
substantial branching ratios into leptons. The difference of sparticle 
masses such as $(m_{\tilde{g}}- m_{\tilde{b}_1})$ 
is determined by measuring
kinematical end points of invariant mass distributions for signal events.

In a previous paper \cite{Hisano:2002xq} we proposed to measure the
edge position of the $m_{tb}$ distribution for the modes (III)$_{1}$
and (IV)$_{11}$, where $m_{tb}$ is the invariant mass of a top-bottom
($tb$) system.  The decay modes are expected to be dominant in the
minimal supergravity model (MSUGRA), since the branching ratios
$\Br(\sbt (\stp) \rightarrow t (b) \tilde{\chi}_1^{\pm})$ could be as
large as 60\%.  We focused on the reconstruction of hadronic decays of
the top quark, because the $m_{tb}$ distribution of the decay makes a
clear ``edge'' in this case.  The parton level $m_{tb}$ distributions
for the modes (III)$_j$ and (IV)$_{ij}$ are expressed as functions of
$\mglu$, $m_{\tilde{t}_1}$, $m_{\tilde{b}_i}$, and the chargino mass
$m_{\tilde{\chi}_j^{\pm}}$: $d\Gamma/dm_{tb}\propto m_{tb}$, and the
edge position (end point) of the $m_{tb}$ distribution $M_{tb}$ for
the modes (III)$_j$ and (IV)$_{ij}$ are written as follows;
\begin{eqnarray}
&&M_{tb}^2({\rm III})_j =  m_t^2
+\frac{m_\stp^2-m^2_{\tilde{\chi}^{\pm}_j}}{2 m_\stp^2}
\left\{
(m_\glu^2-m_{\stp}^2-m_t^2)
\right.
\nonumber\\
&&
\left.
+
\sqrt{
(m_\glu^2-(m_{\stp}-m_t)^2)
(m_\glu^2-(m_{\stp}+m_t)^2)
}
\right\},\nonumber
\label{tb_stop}
\\
&&M_{tb}^2({\rm IV})_{ij}=  m_t^2
+\frac{m_\glu^2-m_{\tilde{b}_i}^2}{2 m_{\tilde{b}_i}^2}
\left\{
(m_{\tilde{b}_i}^2-m_{\tilde{\chi}^{\pm}_j}^2+m_t^2)
\right.
\nonumber\\
&&
\left.
+
\sqrt{
(m_{\tilde{b}_i}^2-(m_{\tilde{\chi}^{\pm}_j}-m_t)^2)
(m_{\tilde{b}_i}^2-(m_{\tilde{\chi}^{\pm}_j}+m_t)^2)
}
\right\}.
\label{tb_sbot}
\end{eqnarray}
The measurement of the end points is sensitive to both
$m_{\tilde{t}_1}$ and $m_{\tilde{b}_1}$, provided that $m_{\tilde{g}}$
and $m_{\tilde{\chi}^{\pm}}$ are determined from other measurements.
The edge height is then closely related to $\sigma(pp\rightarrow
\tilde{g}X)\times {\rm Br(III/IV)}$.  In some model parameters
$M_{tb}({\rm III})_1$ is very close to $M_{tb}({\rm IV})_{11}$.  When
they are experimentally indistinguishable, it is convenient to define
a weighted mean of the end points;
\begin{eqnarray}
\mtbw&=&\frac{\Br({\rm III}) M_{tb}({\rm III})_1+
\Br({\rm IV})_{11}M_{tb}({\rm IV})_{11}}
{\Br({\rm III})+\Br({\rm IV})_{11}},
\cr
\Br({\rm III}) & \equiv & 
\Br({\rm III})_1 +\Br({\rm III})_{11}+ \Br({\rm III})_{21}.
\label{mtbw}
\end{eqnarray}
As the final states $bbW$ from the decay chain  
$\tilde{g}\rightarrow b\tilde{b}_i\rightarrow bW\tilde{t}_1
\rightarrow bbW\tilde{\chi}^{\pm}_1$
(mode (III)$_{i1}$) could have an irreducible 
contribution to the $tb$ final state,  they are included in
the definition of $\mtbw$.

In the previous paper
we showed that the weighted mean $\mtbw$ was successfully measured 
from a fit to the $m_{tb}$ distribution. We also  
discussed the interpretation of the measurement. 
In this paper we significantly extend our study to the 
branching ratio measurements, identification of the stop and 
sbottom decays into heavier charginos, 
and the interpretation in the MSUGRA model.

This paper is organized as follows. In Section~II, we
explain our reconstruction and fitting procedure of
the edge structure in detail. We use a
sideband method to estimate the background distribution due to
misreconstructed events, which plays a key role 
in the edge reconstruction. Monte Carlo simulations show  
that the distribution of the signal modes (III) and (IV)
after subtracting the sideband background
is very close to the parton 
level distribution. In Section~III, we explore the MSUGRA 
parameter space to check if our fitting procedure reproduces 
the end point $\mtbw$ and the number of events going 
through the decay modes (III) 
and (IV). We then turn into 
more delicate issues such as extraction of branching ratios. 
In Section~IV, we discuss the stop and sbottom decays
into the heavier  
chargino $\tilde{\chi}^{\pm}_2$ by looking into the $m_{tb}$ distribution 
of events with 
additional leptons.
In Section~V, we study the MSUGRA parameter region 
where the decay modes (III) and/or (IV) are open, 
and discuss how to extract the fundamental parameters 
using the $m_{tb}$ distributions.
The LHC's potential to 
extract the top polarization arising from the decay
$\tilde{g}\rightarrow
t\tilde{t}_1$ is discussed in Section~VI.  
Section~VII is devoted to 
conclusions. 
\section{Simulation and reconstruction}

In the MSUGRA model  the sparticle spectrum
is parameterized by the universal scalar mass $m_0$, the universal
gaugino mass $M_{0}$, 
the trilinear coupling of the scalar fields
$A_0$ at the GUT scale ($M_{GUT}$), 
the ratio of the vacuum expectation values $\tan\beta$, and sign 
of the higgsino mass parameter $\mu$.
In order to demonstrate the end point reconstruction, we take an MSUGRA
point with $m_0=100$~GeV,  $M_{0}=300$~GeV, $A_0=-300$~GeV, 
$\tan\beta=10$ and $\mu>0$. 
This corresponds to the sample point A1 in Table~\ref{masscross}. 
The masses and
mixings of sparticles are calculated by ISASUSY/ISASUGRA 7.51
\cite{ISAJET}.
Two different event generators, 
PYTHIA 6.1 \cite{PYTHIA} 
and HERWIG 6.4 \cite{HERWIG},
are utilized to generate 
Monte Carlo SUSY events
using the masses and mixings.
In this section we show results with PYTHIA.
The events are then passed through a
fast detector simulation program for the ATLAS experiment, ATLFAST
\cite{ATLFAST}.  
This program performs
jet reconstruction in the calorimeters and
momentum/energy smearing for leptons and photons,
giving a list of reconstructed jets 
as well as identified leptons and photons.
We generate a total of 
$3\times10^6$ SUSY events, which correspond to an integrated
luminosity of 120~fb$^{-1}$. 
In addition to the SUSY events,
$t\bar{t}$ events are generated using PYTHIA
as the standard model background.
The number of generated events  amounts to $1.94 \times 10^8$,
corresponding to an integrated luminosity of 286~fb$^{-1}$.
In the plots in this paper,
the $t\bar{t}$
background is not included unless otherwise noted.

In the detector simulation,
a lepton
is identified if $p_T > 5$~GeV and $|\eta|<2.5$ for an electron, 
and $p_T > 6$~GeV and $|\eta|<2.5$ for a muon, respectively. 
A lepton is regarded as isolated
if it is separated by $ R > 0.4$ from other calorimeter clusters
and the transverse calorimeter energy
in a cone size $R = 0.2$ around the lepton
is less than 10~GeV.
The cone size is 
defined as 
$R = \sqrt{(\Delta \phi)^2 + (\Delta \eta)^2}$,
where $\phi$ and $\eta$ are the azimuthal angle and the pseudo-rapidity,
respectively.

By default,
jets are reconstructed by a cone-based  algorithm with
a cone size $R=0.4$ in the detector simulation 
\footnote{Larger $R$ values, $R = 0.5$ and $0.6$
are also tried to check the dependence on $R$ of the cone-based
algorithm.  For comparison, another jet reconstruction algorithm, the
$K_T$ algorithm, was also tried with $R=0.4, 0.5, 0.6$, and 0.7.  The
comparison is given in the Appendix.}.  After applying the algorithm,
jets having a transverse energy ($E_T$) more than 10~GeV are kept as
reconstructed jets.  The $b$ and $\tau$ tagging efficiencies are set
to be 60\% and 50\%, respectively.  The energy of the reconstructed
jet is recalibrated according to its jet flavor using a
parameterization optimized to give a proper scale of the dijet mass.

\begin{table}[t]
\begin{tabular}{|c|c|c|c|c|c|c|
}
\hline
& $\mglu$&$\mstp$&$\msbt$& $\msbts$ &$m_{\tilde{\chi}^{\pm}}$&$\sigma_{SUSY}$\cr
\hline
A1 & 707& 427 & 570 & 613 &220 &26 \cr
A2 & 706& 496 & 587 & 614&211 &25\cr
T1&707&327&570&613& 220& 30\cr
T2 &707&477&570&612&211& 25\cr
\hline
B & 609& 402 & 504 & 534 &179& 56\cr
C & 931& 636 & 771 & 805 &304& 5\cr
G & 886& 604 & 714 & 763 &285& 7 \cr
I & 831& 571& 648 & 725  &265& 10\cr
\hline
E1      & 515& 273& 521& 634&153& 77 \cr
E2      & 747& 524 & 770& 898&232& 8\cr
\hline
\end{tabular}
\caption{Sparticle masses in GeV and the total SUSY cross
section (${\sigma_{SUSY}}$) in pb for
the parameter points studied in this paper. 
The cross sections are calculated by PYTHIA. }
\label{masscross}
\end{table}

\begin{figure}[ht]
\centerline{\psfig{file=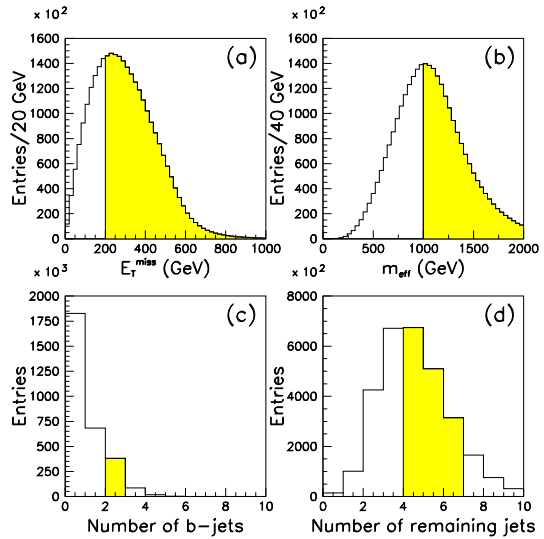,width=3in}} 
\caption{Distributions of 
(a) missing transverse energy,
(b) effective mass,
(c) number of $b$-jets, and
(d) number of remaining jets,
for $3 \times 10^{6}$ SUSY events at the point A1.
Accepted regions are hatched.}
\label{presel}
\end{figure}

First, we apply the following selection cuts 
to the simulated events for the $tb$ signal:
\begin{enumerate}
\item
The missing transverse energy $E_T^{\rm miss}$ is greater than 200~GeV,
where $E_T^{\rm miss}$ is calculated from the reconstructed jets,
leptons, photons, and unreconstructed calorimeter energies.
This calculation is performed before the recalibration of jet energies.
\item 
The effective mass $m_{\rm eff}$ is greater than 1000~GeV,
where $m_{\rm eff}$ is the sum of the
missing transverse energy and transverse momentum of reconstructed
jets, namely 
$m_{\rm eff} = E_T^{\rm miss} + \sum_{all}{p_T^{\rm jet}}$.
\item
There are two and only two
$b$-jets with $p_T > 30$~GeV in an event.
\item
Excluding the two $b$-jets and those identified as tau-jets,
the number of remaining reconstructed jets 
with $p_T > 30$~GeV and
$|\eta|<3.0$ should be between four and six, inclusive.
\end{enumerate}
Distributions of the cut variables are shown in Fig.~\ref{presel}.
The first two cuts are to enhance the SUSY events
against the standard model background events.
The other two cuts are to reduce combinatorial background
(wrong combinations of jets)
in the reconstruction of the top quark.

\begin{figure}[ht]
\centerline{\psfig{file=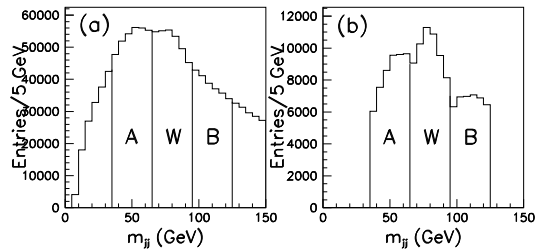,width=3in}} 
\caption{Distributions of the invariant mass of two jets
(a) for all possible combinations, and 
(b) for the combination used to reconstruct a top candidate.
{\bf W} and 
{\bf A}, {\bf B} indicate the $W$ mass region and the $W$ sidebands, 
respectively.
}
\label{mjj}
\end{figure}

\begin{figure}[ht]
\centerline{\psfig{file=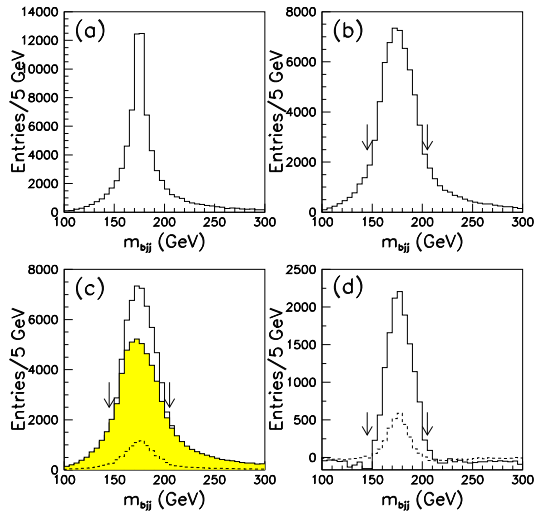,width=3in}} 
\caption{Distributions of 
$m_{bjj}$ (a) before and (b) after the jet energy scaling,
(c) background estimated by the $W$ sidebands
(the hatched histogram) superimposed to the distribution (b),
and 
(d) after subtracting the estimated background.
The mass cut for the top candidate
is indicated by arrows.
Dashed histograms in (c) and (d) show the distributions 
of $t\bar{t}$ events,
where the integrated luminosity of the $t\bar{t}$ events is normalized to
that of the SUSY events.}
\label{mbjj}
\end{figure}

To reconstruct the hadronic decay of the top quark, we  take the
following steps (i)-(iv): 
\begin{itemize}
\item[(i)] 
We first take 
jet pairs consistent with a hadronic $W$ boson decay with a cut on the
jet pair invariant mass $m_{jj}$: $\vert m_{jj}-m_W \vert < 15$~GeV.
The $m_{jj}$ distribution is shown in Fig.~\ref{mjj}(a),
where the selected mass region is marked as {\bf W} (the $W$ mass region).
Although fake $W$ pairs dominate the distribution,
a small bump due to real $W$ bosons can be seen in the mass region.
\item[(ii)] 
The invariant mass of the jet pair and one of the $b$-jets, $m_{bjj}$, is
then calculated.  
All possible combinations of jet pairs and $b$-jets
are tried in an event to select
the combination which minimizes 
the difference $| m_{bjj}-m_t |$.
The distribution of the selected $m_{bjj}$ 
is shown in Fig.~\ref{mbjj}(a). 
\item[(iii)]
The energy and momentum of the jet pair are scaled so
that $m_{jj}=m_W$, and the invariant mass $m_{bjj}$ is recalculated.
The distribution is shown in Fig.~\ref{mbjj}(b). 
The jet
combination is regarded as a top quark candidate 
if $\vert m_{bj\!j}-m_t\vert<30$~GeV.
\end{itemize}

As jets are supplied from gluino or squark decays and
there are several jets in a selected event,
events with a fake $W$ boson 
(a jet pair that accidentally has $m_{jj}\sim m_W$) 
still dominate the $m_{bjj}$ distribution. 
The contribution of the fake $W$ boson in the $W$ mass region
is estimated from the events that contain jet pairs with the
invariant mass in the regions
{\bf A}: $\vert m_{jj}-(m_W-30~{\rm GeV}) \vert < 15$~GeV and 
{\bf B}: $\vert m_{jj}-(m_W+30~{\rm GeV}) \vert < 15$~GeV.
We call them ``the $W$ sidebands''.
The energy and momentum of the jet pairs
are then scaled linearly to be in the $W$ mass region 
$\vert m_{jj}-m_W\vert<15$~GeV. 
The $m_{jj}$ distributions before the linear scaling 
are shown for the $W$ region and the $W$ sidebands in Fig.~\ref{mjj}(b).
The distribution of the fake top quark candidates is
estimated by following the same steps (ii)-(iv) for the 
scaled jet pair (see the hatched histogram in Fig.~\ref{mbjj}(c)). 
The ``true'' distribution is obtained 
by subtracting the background distribution estimated by the $W$ sidebands
from the original distribution (see Fig.~\ref{mbjj}(d)).
The estimation is based on an assumption that the background comes from
the jets without significant correlation with the $b$-jets. 

\begin{figure}[ht]
\centerline{\psfig{file=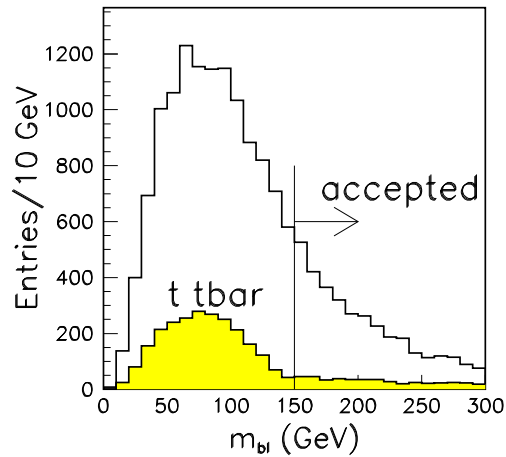,width=3in}} 
\caption{Distribution of minimum $m_{bl}$
for events having a top candidate.
Open and hatched histograms are for SUSY events and $t\bar{t}$
events, respectively. The loose lepton cut is also shown.}
\label{mbl}
\end{figure}

\begin{figure}[ht]
\centerline{\psfig{file=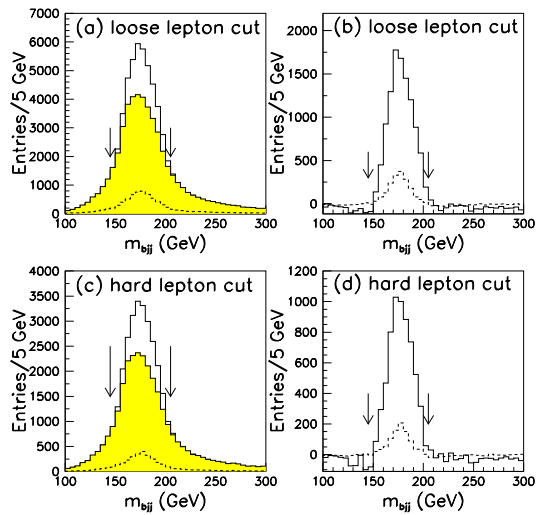,width=3in}} 
\caption{
(a) and (b): Distributions of $m_{bjj}$
with the loose lepton cut 
before and after sideband subtraction, respectively.
(c) and (d) : Distributions of $m_{bjj}$
with the hard lepton cut 
before and after the sideband subtraction.
The meaning of the histograms is as same as in Fig.~\ref{mbjj}(c) and (d).
}

\label{mbjj1}
\end{figure}

The $t\bar{t}$ production is the dominant standard model background.
The $m_{bjj}$ distributions of the $t\bar{t}$ events
surviving at this stage are shown in Fig.~\ref{mbjj}(c) and (d) 
before and after the sideband background subtraction, respectively.
The integrated
luminosity of the $t\bar{t}$ events is normalized to that of the
SUSY events.
To reduce the $t\bar{t}$ events,
one of the following two lepton cuts may be used, 
depending on the signal/background situation.
\begin{itemize}
\item
Loose lepton cut:
If there are high-$p_T$ isolated leptons,
the invariant mass of any high-$p_T$ lepton and the remaining $b$-jet
($m_{bl}$)
should be greater than 150~GeV.
The distribution of the minimum $m_{bl}$ in events with a top candidate
is shown in Fig.~\ref{mbl}. 
\item
Hard lepton cut:
An event should have no isolated leptons.
\end{itemize}
The cuts reduce the fraction of the events having a $t\bar{t}$ pair,
where one of the top quarks decays leptonically.
The $m_{bjj}$ distributions with the lepton cuts
are shown in Fig.~\ref{mbjj1}.
The numbers of selected events are listed in Table~\ref{lepton}. 
Hereafter we use the loose lepton cut.

\begin{table}[ht]
\begin{tabular}{|l|rr|rr|}
\hline
& \multicolumn{2}{c|}{SUSY point A1} & \multicolumn{2}{c|}{$t\bar{t}$} \\
sideband subtraction & before & after & before & after \\
\hline
No lepton cut    & 59174 & 13340 & 8168 & 2764 \\
                 &       &       & (0.138) & (0.207) \\
\hline
Loose lepton cut & 47171 & 10487 & 5789 & 1777 \\
                 &       &       & (0.123) & (0.169) \\
\hline
Hard lepton cut  & 26915 &  6114 & 2671 &  884 \\
                 &       &       & (0.099) & (0.145) \\
\hline
\end{tabular}
\caption{Numbers of events having a top candidate
with or without the lepton cuts.
The integrated luminosity is 120~fb$^{-1}$.
The numbers in parentheses are the ratios 
of the $t\bar{t}$ events to the SUSY events.
}
\label{lepton}
\end{table}

\begin{figure}[t]
\centerline{\psfig{file=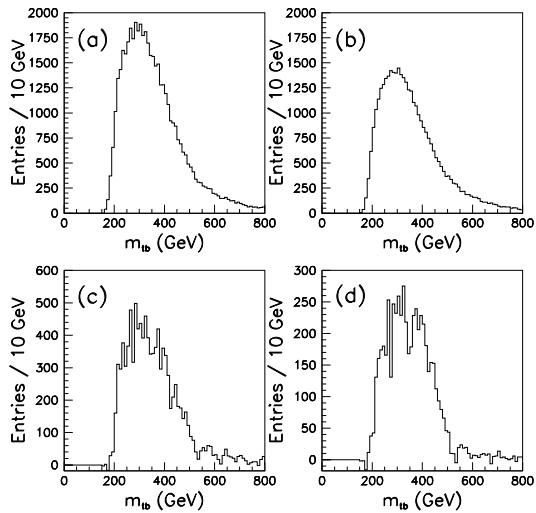,width=3in}}
\caption{
(a) Signal $m_{tb}$ distribution for the sample point A1 in
Table~\ref{masscross}, 
(b) the estimated background distribution from the sideband
events,  (c) (a)$-$(b), and (d) the $m_{tb}$ distribution for the modes 
(III)$_1$
and (IV)$_{11}$ in Eq.~(\ref{gluinodecay}), and the decay modes 
(III)$_{i1}$ irreducible to the mode (III)$_1$.
}
\label{mtb}
\end{figure}

For the remaining events,
the top candidate is then combined with the other $b$-jet,
which is not used to reconstruct the top candidate,
to calculate the invariant mass of the $tb$ system $m_{tb}$.
The distribution is shown in Fig.~\ref{mtb}(a).
However, the expected $tb$ end point is not clearly visible 
in the $m_{tb}$ distribution due to the fake $W$ events.
Here we can again utilize the $W$ sidebands for the background estimation.
The estimated background distribution is shown in Fig.~\ref{mtb}(b), 
which is
obtained by averaging distributions from the sidebands {\bf A} and {\bf B}. 
The estimated background distribution is subtracted from the
signal distribution in Fig.~\ref{mtb}(c).  The {\it corrected} signal
distribution (c) shows a better end point structure
compared to (a). 

By using the information from the event generator, one can show the 
sideband method is indeed reproduces the $m_{tb}$ distribution 
for the signal modes (III) and (IV). 
Fig.~\ref{mtb}(d) is the same distribution 
as Fig.~\ref{mtb}(c) but for the events
which contain either the decay chain  (III)$_1$, 
the  decay chain  (IV)$_{11}$, or the decay chains (III)$_{i1}$,
which are 
irreducible to the chain (III)$_1$; $\glu\rightarrow b\sbti 
 \rightarrow bW\stp
\rightarrow bbW\chapm$. Note that if $bW$ has an invariant mass
consistent to a top quark, the decay is kinematically equivalent to
the mode (III)$_1$.  Fig.~\ref{mtb}(d) shows two end points 
as expected ($M_{tb}({\mathrm{III}})_1=471$~GeV and
$M_{tb}({\mathrm{IV}})_{11}=420$~GeV),  demonstrating that the
sideband method works well. 

Note that the signals from the modes (III) and (IV) in
Eq.~(\ref{gluinodecay}) are significant in the total selected
events. The total distribution shown in
Fig.~\ref{mtb}(c) contains  contribution from
mis-reconstructed events  such as other gluino decay chains with $W$ and 
$b$-jets, 
or stop/sbottom pair productions. 

We fit the total distribution shown in Fig.~\ref{mtb}(c),
which is made with a bin size $\Delta m =$ 10~GeV,
by a simple
fitting function described with
the end point
$M_{tb}^{\rm fit}$, the edge height $h$ per $\Delta m $ bin, 
and the smearing parameter
$\sigma$ originated from the jet energy resolution;
\begin{eqnarray}
f(\mtb)\!&=&\!\frac{h}{\mtbfit } \int_{m_t + m_b}^{\mtbfit}
\!\frac{m}{\sqrt{2\pi}\sigma} 
\exp\left({ -\frac{1}{2} \left[\frac{m-\mtb}{\sigma} \right] ^2 }\right) dm
. \cr&&
\end{eqnarray}
To reduce the number of free parameters for good convergence
capability of the fit, we set the smearing parameter $\sigma$ to be
10~\% of the end point $M_{tb}^{\rm fit}$.  This assumption is based
on the dijet mass resolution of the ATLAS detector~\cite{TDR}.  We
assume that the signal distribution is sitting on a
linearly-decreasing background expressed by a function $a (m_{tb} -
\mtbfit) + b$.  The parameters $a$ and $b$ are also determined by the
fit.

\begin{figure}[t]
\centerline{\psfig{file=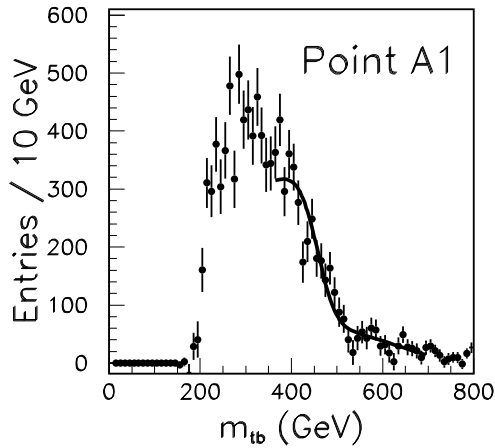,width=3in}}
\caption{A fit to the $m_{tb}$ distribution at the  point A1.}
\label{tblfit}
\end{figure}

The fit results, especially the edge height $h$, depend on the mass
range used for the fit.  We therefore apply an iterative fitting
procedure to obtain stable results.  The initial $\mtbfit$ value is
determined by fitting a rather wide mass range $350$~GeV$< m_{tb}
<~$800~GeV (45 histogram bins).  In the next step, lower 8 bins and
higher 25 bins with respect to the $\mtbfit$ value are used to obtain
new fit results including the new $\mtbfit$ value.  The relatively
small number of lower bins is chosen because the edge height is
sensitive to the distribution in the mass region near the end point.
Including too many lower bins is found to degrade the sensitivity.
The higher bins determines the linearly-decreasing background.  This
step is repeated until the fit results become stable.  The number of
iterations is typically three or four.  If the edge height $h$ has a
large fit error, the number of lower bins is increased to 9 or 10,
until the error becomes reasonably small (typically less than 15~\%).
The fit result is shown in Fig.~\ref{tblfit}.  We obtain $M^{\rm
fit}_{tb}=455.2 \pm 8.2$~GeV and $h = 271 \pm 23$~/(10~GeV), where the
errors are statistical.

\section{The $m_{tb}$ distributions in the SUSY model points}

In the previous section we discuss the selection of the $tb$ events
and the extraction of the edge structure in the $m_{tb}$ distribution
originated from the gluino decays (III)$_1$, (III)$_{i1}$ or
(IV)$_{i1}$.  Two values are obtained from the $m_{tb}$ distribution;
the end point $M^{\rm fit}_{tb}$ is directly related to the stop,
sbottom, and gluino masses, while the edge height $h$ is sensitive to
the gluino production cross section and the decay branching ratios.

This section is aimed to compare the reconstructed values with
expectations.  We will see that the kinematical quantities $M^{\rm
fit}_{tb}$ and the peak value of the effective mass distributions
($M_{\rm eff}$) agree very well to $\mtbw$ and $m_{\tilde{q}}+
m_{\tilde{g}}$, respectively.  On the other hand, the reconstructed
edge height $h$, which is related to the total number of gluinos
decaying through the modes (III) or (IV), is dependent on the event
generator, and also on the decay patterns of the neutralino, chargino
and squarks.  We clarify key issues to reduce the uncertainties in
this section.

\subsection{Model parameters}

The gluino decay widths and the decay kinematics depend on the squark
masses and mixings. The charginos and neutralinos in the decay chains
(I)-(IV) further decay into jets/leptons, and the decay patterns
depend on the electroweak SUSY parameters such as
$\mu,M_1,M_2,\tan\beta$ and the slepton masses.  The $m_{tb}$
distribution is therefore a function of all SUSY parameters.  To check
the validity of our reconstruction and subtraction scheme, we study
the $m_{tb}$ distributions for various sets of SUSY parameters.
    
The reference point A1 introduced in the previous section corresponds
to the MSUGRA model with a large negative $A_0$ value.  The stop and
sbottom masses can be changed by varying the $A_0$ value without
changing other sparticle masses drastically (see Section~V for the
detail).  We thus make another MSUGRA point A2, by changing only the
$A_0$ value from $-300$~GeV to $300$~GeV from the point A1.  In
addition, two non-MSUGRA points T1 and T2 are made from the point A1,
where the chargino/neutralino sector and the gluino mass are kept
unchanged but the stop masses and mixing are modified by changing the
mass parameter $m_{\tilde{t}_R}$.  The stop mass parameters and the
mixing angle at the low energy scale are listed in
Table~\ref{stopsector}.

\begin{table}
\begin{center}
\begin{tabular}{|c||c|c|c|c|c|}
\hline
&$m_{\tilde{t}_1}$& $m_{\tilde{t}_R}$  & $m_{\tilde{t}_L}$&
 $A_t$&$\theta_{\tilde{t}}$\cr
\hline
A1 &427 & 482 & 573 & $-655$ & 0.99  \cr
A2 &496 & 521 & 591 & $-457$ & 1.01  \cr
T1 &327 & 366 & 573 & $-655$ & 1.14  \cr
T2 &477 & 551 & 573 & $-655$ & 0.84  \cr
\hline
\end{tabular}
\end{center}
\caption{The lighter stop mass and the  relevant SUSY mass parameters in GeV 
and the stop mixing angles for the points
A1, A2, T1 and T2.}
\label{stopsector}
\end{table}

Next we choose MSUGRA points B, C, G and I, which were discussed 
in the paper 
\cite{ellis}.
They satisfy $m_{\tilde{g}}>m_{\tilde{t}_1}+m_t$ so that the decay
mode (III)$_1$ is open.  The points also satisfy the condition
$m_{\tilde{g}} <~1$~TeV, which is required for a statistical reason.
The gluino production is dominated by the process
$qg\rightarrow\tilde{q}\tilde{g}$ in the case of $m_{\tilde{q}} \sim
m_{\tilde{g}}$, and the production cross section is parameterized as
\cite{Hisano:2002tk}
\begin{equation}
\sigma(\tilde{q}\tilde{g})+\sigma(\tilde{q}^{*}\tilde{g}) = 1.74 \times 
\left(\frac{m_{\tilde{g}}}{\rm TeV}\right)^{-5.78} ({\rm pb}).
\end{equation}
For example the cross section is 14~pb and 3.1~pb for
$m_{\tilde{g}}=700$~GeV and 900~GeV, respectively.  Assuming that the
branching ratios and the reconstruction efficiency are equal to those
of the point A1, the gluino mass must be less than 1~TeV to
reconstruct more than 2000 SUSY $tb$ events after the sideband
subtraction for an integrated luminosity of 100~fb$^{-1}$.

Finally we select two more points E1 and E2, where only the decay
chain (III)$_1$ is kinematically open and the other two body gluino
decays into squarks are closed.  This happens when $m_0$ and $A_0$ are
large.  As the SUSY events at these points typically contain four
bottom quarks, the combinatorial background for the $tb$
reconstruction is significantly large.

\begin{table}
\begin{tabular}{|c||c|c|c|c|c|}
\hline
&$M_0$& $m_0({\tilde{q}})$& $m_0({H})$& $A_0$ & $\tan\beta$ \cr 
\hline
\hline
A1&300& 100 & 100 &$-300$& 10 \cr
A2&300& 100 & 100 &300 &10\cr
\hline
B&255& 102 & 102 & 0  &10\cr
C&408 & 92 & 92 & 0 &10 \cr
G&383 & 125 & 125 & 0& 20\cr
I&358 & 188 & 188 & 0& 35 \cr
\hline
E1&200 & 500 & 200 & $-1000$ & 10 \cr
E1&300 & 700 & 500 & $-1000$ & 10\cr
\hline
\end{tabular}
\caption{The universal parameters at the GUT scale for the points 
we study. Units are in GeV except $\tan\beta$. }
\label{gut}
\end{table}

For the points we study in this paper, we list the relevant sparticle
masses and the universal parameters at the GUT scale in
Table~\ref{masscross} and Table~\ref{gut}, respectively.  At each SUSY
point we generate two Monte Carlo data samples, each having $3\times
10^6$ events, where one is generated by PYTHIA and the other by
HERWIG.  The cross sections are summarized in Table~\ref{masscross}.
The SUSY cross section for the point C is the smallest among them,
where the generated $3\times 10^6$ events correspond to an integrated
luminosity of 600 fb$^{-1}$.

\subsection{Reconstruction of kinematic variables}
\label{kinematical}
The fit results of the $m_{tb}$ distributions are listed in
Table~\ref{fittable}, where we follow the fitting procedure described
in the previous section.  The weighted end point $M^{\rm w}_{tb}$ is
defined in Eq.~(\ref{mtbw}) and the relevant branching ratios are
listed in Table~\ref{branch}.  In Table~\ref{fittable}, $\nall$ is the
number of $tb$ events after the sideband subtraction, while $\nedge$
is the number of $tb$ events after the subtraction, with one and only
one gluino decaying through the mode (III)$_1$, (III)$_{i1}$ or
(IV)$_{i1}$.  Generator information is used to obtain $\nedge$.

\begin{figure}[ht]
\centerline{\psfig{file=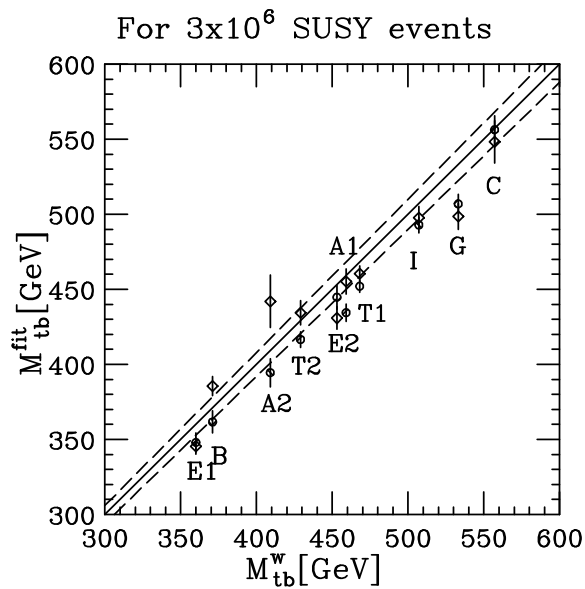,width=3in}} 
\caption{ 
Relation between 
$M^{\rm w}_{tb}$ and 
$M^{\rm fit}_{tb}$ for the sample points. The solid line corresponds to  
$M^{\rm w}_{tb}= M^{\rm fit}_{tb}$ and dashed lines 
to $M^{\rm w}_{tb}(1\pm 0.02)= M^{\rm fit}_{tb}$. Bars with a diamond
and a circle correspond to  PYTHIA  and HERWIG samples, respectively. 
}
\label{endpoint}
\end{figure}

The relation between $M_{tb}^{\rm fit}$ and $M_{tb}^{\rm w}$ is shown
in Fig.~\ref{endpoint}.  The fitted value $M^{\rm fit}_{tb}$ increases
linearly with the weighted end point $M^{\rm w}_{tb}$ in the mass
range $350$~GeV$<M^{\rm w}_{tb}<~600$~GeV, and tends to be lower than
$M^{\rm w}_{tb}$.  In the previous
literature~\cite{Hinchliffe:1996iu,TDR}, the lower mass value was
understood as the effect of particles missed outside the jet cones.
In the literature invariant mass distributions of the events with same
flavor and opposite sign leptons are studied, which comes from the
cascade decay of the squarks $\tilde{q}\rightarrow
q\tilde{\chi}^0_2\rightarrow ql \tilde{l}\rightarrow qll
\tilde{\chi}^0_1$.  The end point of the $m_{jll}$ distribution is
lower than the parton level $qll$ end point by about 10\%.

In the study of the squark cascade decay $ \tilde{q}\rightarrow
qll\tilde{\chi}_1^0$, the end point agrees better with the parton level
one by changing the jet cone size to $R=0.7$.  In our study at the
reference point A1, the reconstructed number of events is
significantly reduced for $R=0.7$.  We find that using the jet cone
size $R=0.5$ leads to a better reconstruction and a larger
$M_{tb}^{\rm fit}$, which is closer to $\mtbw$.  Such comparison might
be useful to estimate the true end point. We discuss the dependence on
the jet finding algorithm and the cone parameters in Appendix.

We note that our definition of $M^{\rm w}_{tb}$ might be too simple if
the reconstruction efficiencies of the decay modes (III) and (IV) are
very different.  In addition, the weighted average should not be
applied when $M_{tb}({\rm III})_1$ and $M_{tb}({\rm IV})_{11}$ differ
by more than 80~GeV, as we use the events with $m_{tb}>\mtbfit-80$~GeV
for the fit.

\begin{table}[ht]
\begin{tabular}{|c||cccc|cc|}
\hline
	&gen	&$\mtbw$[GeV]	&$M^{\rm fit}_{tb}$[GeV] & $h$/(10GeV) & $\nedge$ & $\nall$  \cr
\hline
A1	&PY	&459	&455.2 $\pm$ 8.3	&271.4$\pm$22.7	& 5846 & 10487\cr
	&HW	&	&434.5 $\pm$ 5.8	&354.8$\pm$23.3	& 6685 & 11470\cr
\hline
A2	&PY	&409	&442.0 $\pm$17.5	&153.0$\pm$20.6	& 3064 & 7525 \cr
	&HW     &       &394.4 $\pm$9.5  &190.6$\pm$21.8 & 3095 & 7805 \cr
\hline
T1      &PY     &468    &460.3 $\pm$ 5.4 &327.0 $\pm$ 21.6 & 6620 & 11659 \cr
	&HW	&	&452.0 $\pm$ 3.9&447.5 $\pm$ 23.5& 8170 & 14050 \cr
\hline
T2	&PY	&429	&434.5 $\pm$ 8.1&223.2 $\pm$ 21.6& 4461 & 8466 \cr
	&HW     &       &416.6 $\pm$ 5.2 &321.0 $\pm$ 23.2 & 5378 & 9592 \cr
\hline
B	&PY	&371	&385.6 $\pm$ 6.3&226.9 $\pm$ 19.6& 2801 & 5396 \cr
	&HW	&	&361.7 $\pm$ 7.3&223.5 $\pm$ 21.3& 3105 & 5935 \cr
\hline
C	&PY	&557	&548.2 $\pm$14.1&142.4 $\pm$ 17.5& 4026 & 11228 \cr
	&HW	&	&556.3 $\pm$9.4  &178.2 $\pm$ 17.9& 4395 & 12704 \cr
\hline
G	&PY	&533	&498.5 $\pm$ 8.6&244.1 $\pm$ 23.8& 5784 & 13630 \cr
	&HW	&	&506.9 $\pm$ 6.4&325.5 $\pm$ 22.9& 6248 & 15039 \cr
\hline
I	&PY	&507	&497.8 $\pm$ 7.3&289.7 $\pm$ 24.0& 6016 & 13752 \cr
	&HW	&	&492.9 $\pm$ 5.2&383.9 $\pm$ 24.5& 6661 & 14968 \cr
\hline
E1	&PY	&360	&345.5 $\pm$ 5.3&270.6 $\pm$ 23.7& 2778 & 4595\cr
	&HW	&	&348.0 $\pm$ 6.3&251.5 $\pm$ 23.9& 3169 & 5167\cr
\hline
E2	&PY	&453	&430.8 $\pm$ 7.5&352.0 $\pm$ 33.9& 5577 & 16490\cr
	&HW	&	&444.7 $\pm$ 8.0&324.6 $\pm$30.9& 4394 & 15724\cr
\hline
\end{tabular}
\caption{Fit results of the $m_{tb}$ distributions and 
numbers of $tb$ events after the sideband subtraction 
$\nedge$ and $\nall$ for $3\times 10^6$ 
Monte Carlo events. 
The fit does not include the $t\bar{t}$ background, and 
the loose lepton cut is applied. 
$\nall$ is the total number of $tb$ events, while 
$\nedge$ is the number of events with one and only one 
gluino decays into mode (III)$_1$, (III)$_{i1}$ or (IV)$_{i1}$.
``PY'' is for  PYTHIA and ``HW'' is for HERWIG.}
\label{fittable}
\end{table}

\begin{table}
\begin{tabular}{|c||c|c|c|c|c|c|c|}
\hline
&        (III)$_1$&  (IV)$_{11}$ & (IV)$_{21}$ & (III)$_{11}$ & (III)$_{21}$& sum &  $bbX$  \cr 
\hline
A1  &  11.0 &  6.7  &  1.4  &  3.4 &   2.7 &   25.3 & 43.4 \cr
A2  &   3.1 &  6.5  &  1.6  &  1.4  &  0.4 &   13.1 & 32.0 \cr
T1  &  24.5 &  3.2  &  0.8  &  5.0 &   3.0 &   36.5 & 56.3\cr
T2  &   4.3 &  9.9  &  2.2  &  0.5 &   2.1 &   19.0 & 36.2\cr
B &   4.1  &  8.2 &   2.3 &   0.9 &   1.7  &  17.3 & 33.5\cr
C  &  7.2  &  5.3 &   1.3 &   0.9 &   0.8  &  15.4 & 38.5 \cr
G  &  6.6  &  7.5 &   1.2 &   0.5 &   0.8  &  16.6 & 40.4 \cr
I  &  6.2  & 11.1 &   0.7  &  0.0 &   0.7  &  18.7 & 47.3\cr
E1  &  78.5  & 0 &  0   & 0  &  0   &  78.5 & 99\cr
E2  &  42.6 & 0 &  0   & 0  &  0   &  42.6 & 98\cr
\hline
\end{tabular}
\caption{Branching ratios of gluino cascade decays in \%. 
The decay modes (III) and (IV) are defined in Eq.~(\ref{gluinodecay}).
The ``sum'' is total of all (III)$_1$, (III)$_{i1}$ and (IV)$_{i1}$ 
decay modes.
The `` $bbX$'' 
is the branching ratio of the gluino decaying
into $\tilde{t}_i$ or $\tilde{b}_i$, so that 
two bottom quarks appear in the decay products.}
\label{branch}
\end{table}

\begin{figure}[ht]
\centerline{\psfig{file=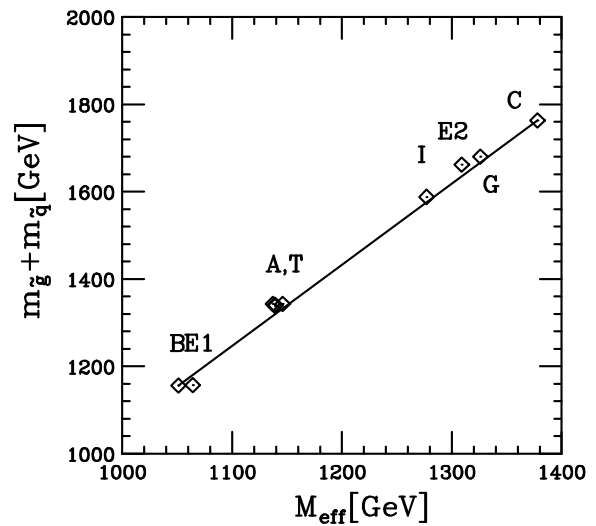,width=3in}}
\caption{ Relation between  $M_{\rm eff}$ and 
$m_{\tilde{g}}+m_{\tilde{q}}$  
for the sample points in Table~\ref{masscross}
(HERWIG samples). 
The line  shows a linear 
fit of $m_{\tilde{g}} +m_{\tilde{q}}$ as a function of 
$M_{\rm eff}$. 
}
\label{ameff}
\end{figure}

We now discuss the relation between  $M_{\rm eff}$ and 
$m_{\tilde{g}}, m_{\tilde{q}}$ 
using the $tb$ samples. 
We find that 
sum of the masses 
$m_{\tilde{g}}+m_{\tilde{q}}$ has a linear dependence on  
$M_{\rm eff}$ as shown in Fig.~\ref{ameff}.  
The deviation of the sample points
from a linear fit is less than 5\%.
The plot is for HERWIG samples  
and we find that PYTHIA and HERWIG give consistent results
for the relation between  $M_{\rm eff}$ and the masses.

The $tb$ sample contains two and only two $b$-jets originating from a
gluino decay, therefore the $m_{\tilde{g}}$ dependence of $M_{\rm
eff}$ is expected.  One may wonder that the $\tilde{g}\tilde{g}$
production might affect the $M_{\rm eff}$ value.  This is, however,
not the case for the points we study.  Indeed, the points E1 and E2,
where $m_{\tilde q}\gg m_{\tilde{g}}$, also satisfy the linear
relation between $M_{\rm eff}$ and $m_{\tilde{q}}+m_{\tilde{g}}$.
Note that the quark parton distribution is harder than the gluon
parton distribution.  Therefore $\sigma(pp\rightarrow
\tilde{q}\tilde{g})\gg
\sigma(pp\rightarrow \tilde{g}\tilde{g})$ 
in  a wide MSSM parameter region. This is why  
$M_{\rm eff}$ becomes a very good function of
$m_{\tilde{g}}+m_{\tilde{q}}$.

In Ref.~\cite{tovey} the relation between $M_{\rm eff}$ and the
effective SUSY scale is studied, where the effective SUSY scale is
defined as the cross section weighted mean of the masses of two
sparticles initially produced in $pp$ collisions.  On the other hand,
we actively select the $\tilde{g}\tilde{q}$ productions by requiring
two tagged $b$-jets.  This leads to a clear dependence of $M_{\rm
eff}$ and $m_{\tilde{q}}+m_{\tilde{g}}$.

\subsection{Number of $tb$ events}

We now discuss the relation between the edge height $h$, the number of
reconstructed $tb$ events, and the reconstruction efficiencies.  The
total number of the ``edge'' events $\nedge$ arising from the decay
chains (III) and (IV) may be estimated from $M_{tb}^{\rm fit}$, $h$
per the bin size $\Delta m$ as follows,
\begin{equation}
\nedge   \sim \nfit = 
\frac{h}{2}\left(\frac{m_t}{M^{\rm fit}_{tb}}  + 1\right)\times 
\frac{M^{\rm fit}_{tb}-m_{t}}{\Delta m }.
\label{nsig}\end{equation}
This formula is obtained by assuming the parton level distribution,
and equating the minimum of the $m_{tb}$ distribution from the decay
chain (III) or (IV) to $m_t$. This is a good approximation for
reasonable SUSY parameters.

\begin{figure}[ht]
\centerline{\psfig{file=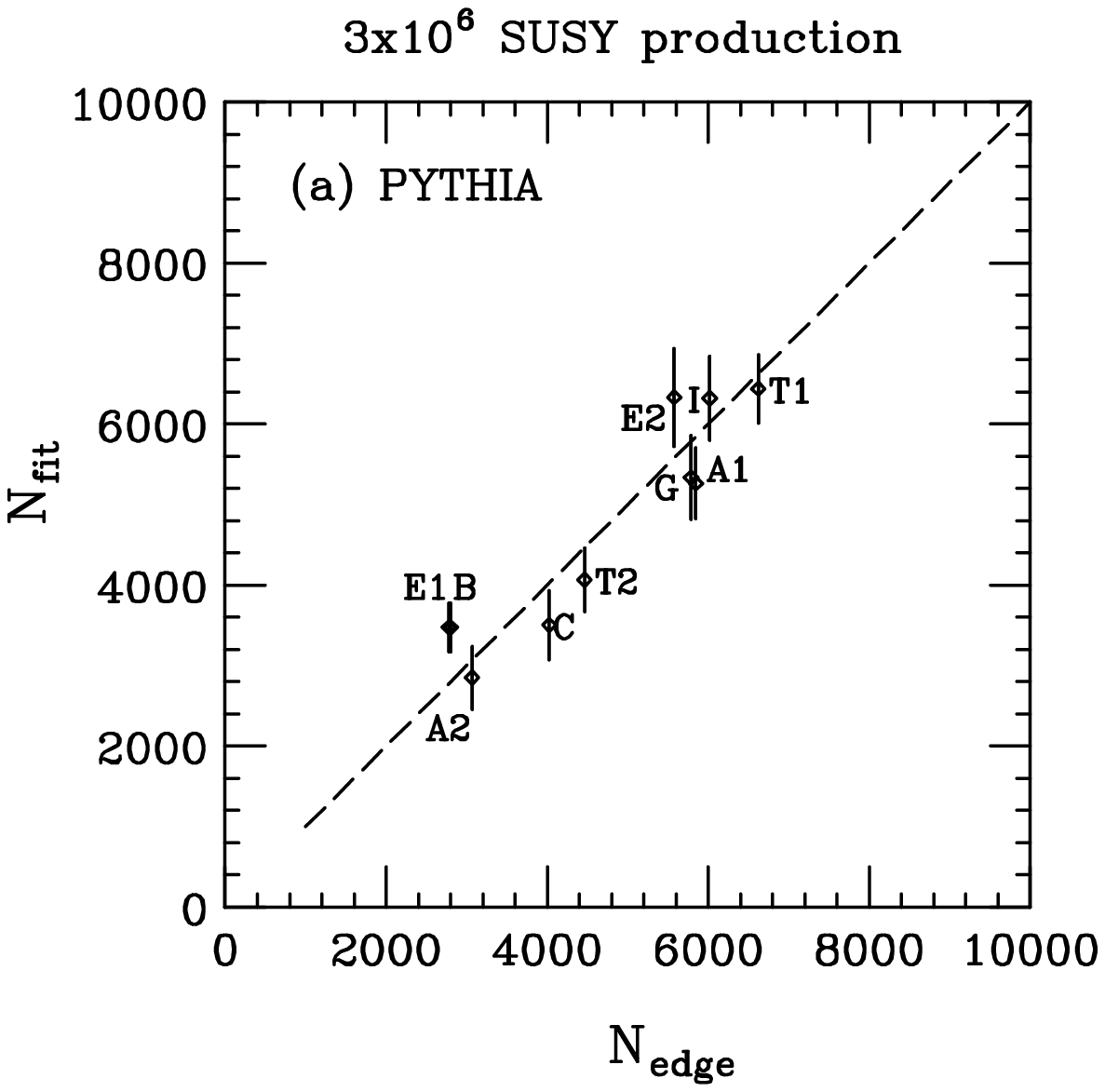,width=2.5in}} 
\centerline{\psfig{file=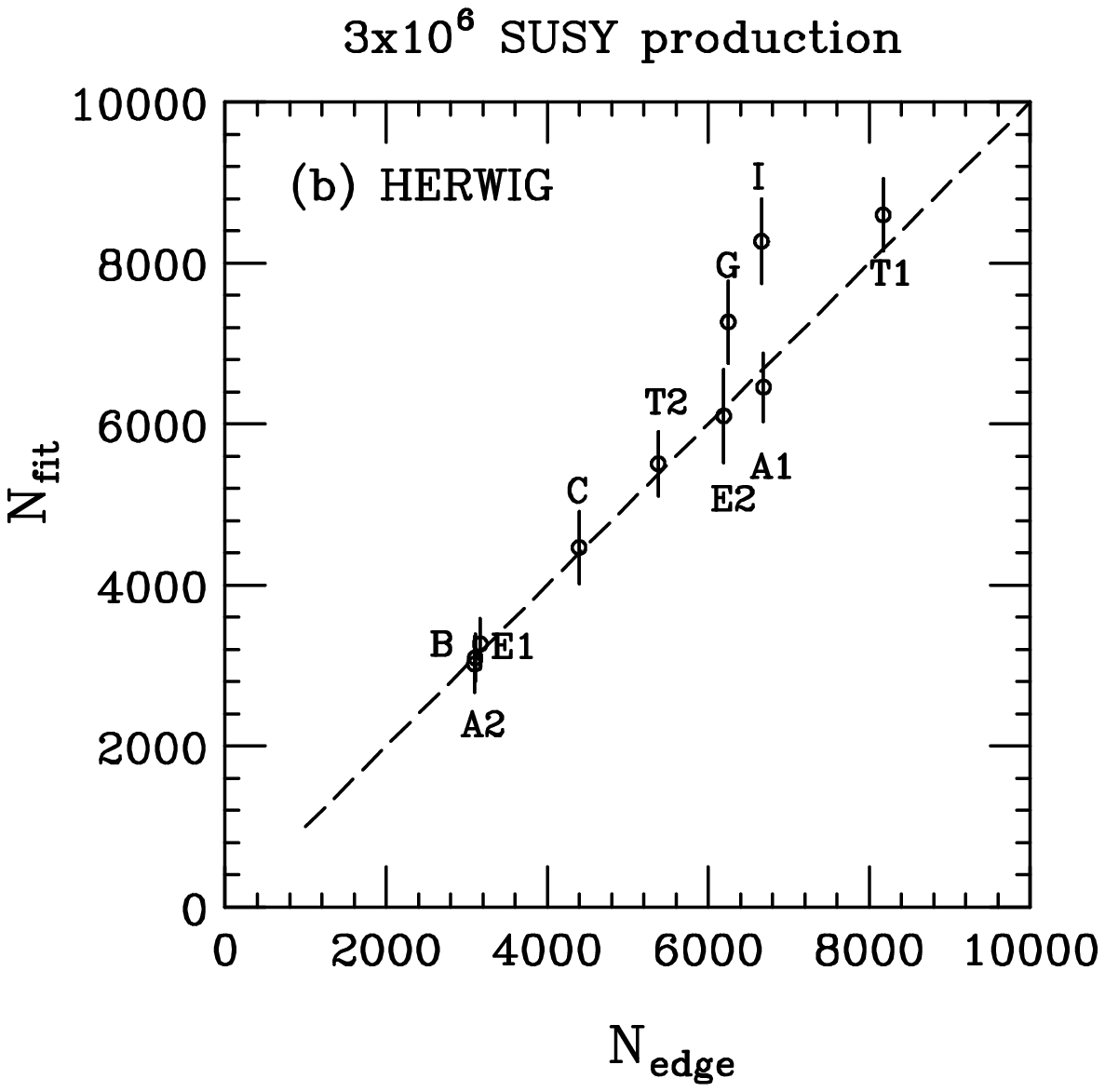,width=2.5in}} 
\centerline{\psfig{file=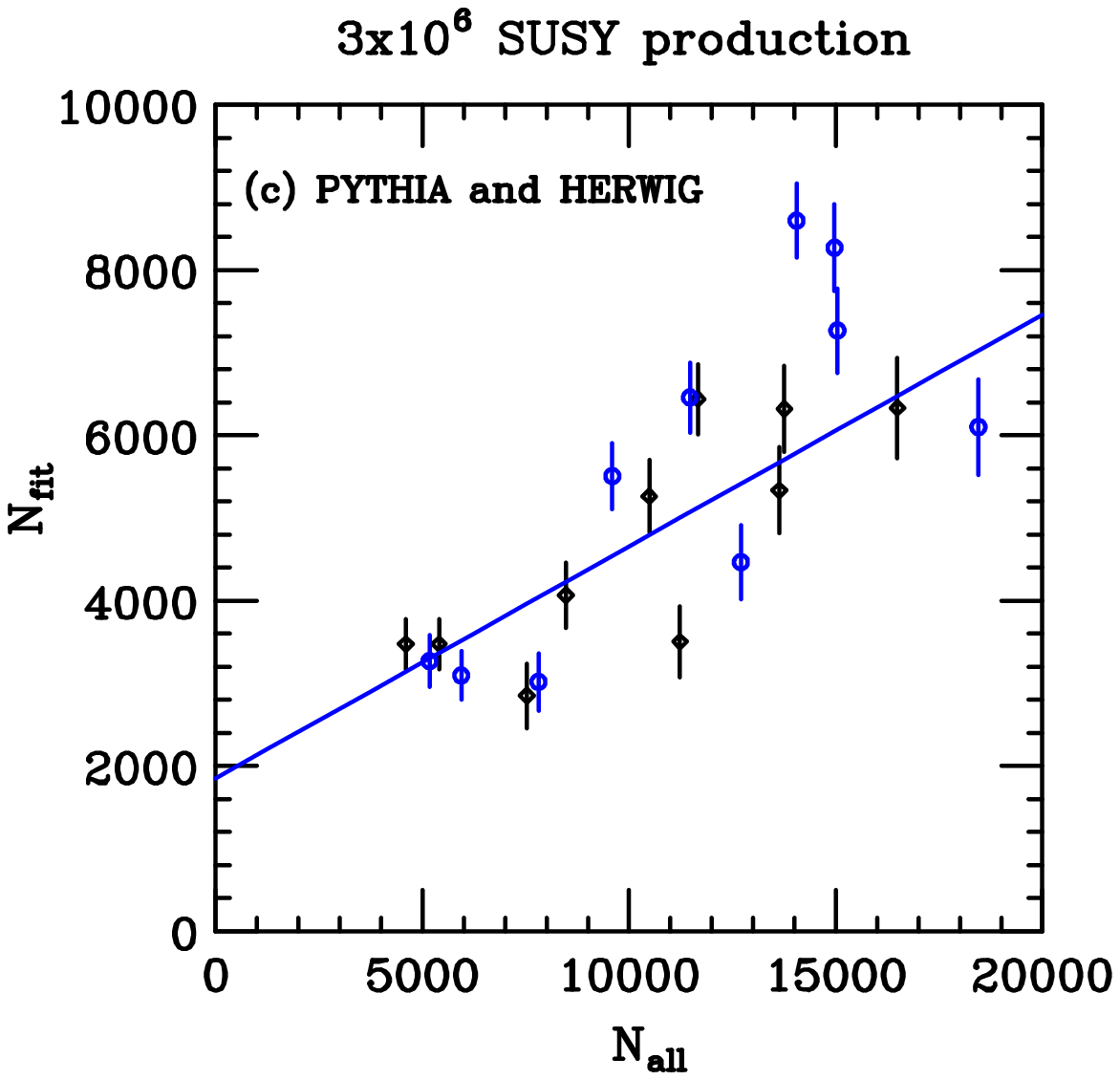,width=2.5in}} 
\caption{Relations between $\nedge$ and $\nfit$ for (a) PYTHIA and
(b) HERWIG. (c) Relation between $\nall$ and $\nfit$.
The line in (a) and (b)  shows $\nfit=\nedge$,
while the line in (c) is a result of linear fit.
}
\label{reconstruct}
\end{figure}

In Fig.~\ref{reconstruct}(a) and (b) we compare $\nedge$ in
Table~\ref{fittable} with $\nfit$.  We find a very good agreement
between them both for the PYTHIA and HERWIG samples.  On the other
hand, the correlation between $\nall$ and $\nfit$ is much worse as
shown in Fig.~\ref{reconstruct}(c).  The number $\nall$ receives
contribution from other gluino cascade decay chains such as the modes
(I) and (II), as well as contributions from the stop and the sbottom
pair productions.  By the end point fit, we extract number of events
coming from only the modes (III)$_{1}$, (III)$_{i1}$ and (IV)$_{i1}$.

$\nedge$, and then $\nfit$, must be related to the number of gluino
decays through the cascade decay chains (III) and (IV) via the
reconstruction efficiencies. For the points we study, the number of
produced $\tilde{g}\tilde{g}$ events, $N(\tilde{g}\tilde{g})$, is
typically 10\% to 14\% of the total SUSY production events, while the
number of $\tilde{q}^{*}\tilde{g}$ and $\tilde{q}\tilde{g}$ production
events $N(\tilde{q}^{*} \tilde{g})+ N(\tilde{q} \tilde{g})$ ranges
from 42\% to 51\%.  The gluino decay branchings ratios are listed in
Table~\ref{branch}.  The number of events $\nprod$, where one gluino
decays through the modes (III) or (IV) and the other squark or gluino
decay does not produce any bottom quark, is given as follows;
\begin{eqnarray}\label{nprod}
\nprod&=&2 N(\tilde{g}\tilde{g}) \left(1-\Br(\tilde{g}\rightarrow bbX) \right)
\Bredge\cr
\rm 
&&+\left(N(\tilde{g}\tilde{q}) + N(\tilde{g}\tilde{q}^*)\right) \Bredge,
\cr
\rm 
Br(edge)&\equiv & \rm Br(III)_1 + Br(III)_{11}+ Br(III)_{21} \nonumber \\
        &       &\rm + Br(IV)_{11},
\end{eqnarray}
where $\Br(\tilde{g}\to bbX)$ is the branching ratio of the gluino
decaying into stop or sbottom, thus having two bottom quarks in the
final state.  The reconstruction efficiency of the $tb$ edge mode,
${\epsilon}_{tb}$, is given as
\begin{equation}
\epsilon_{tb}={\nedge / \nprod}. 
\end{equation}
If the efficiency does not strongly depend on uncertainty in
hadronization and the model parameter dependence can be corrected from
other measurements, we can extract the number $\nprod$ from the
experimental data.

The major uncertainty in hadronization may be estimated by the
generator dependence of the reconstruction efficiency $\epsilon_{tb}$.
The edge height $h$ for the HERWIG sample is significantly larger than
that of the PYTHIA sample in Table~\ref{fittable}, except the points
E1 and E2, and the difference is more than 20\% at many points.  We
note the difference is small before we apply the sideband subtraction.
For example, the numbers of $tb$ events before and after the sideband
subtraction are 9695 (10180) and 2462 (2949) for the PYTHIA (HERWIG)
sample, respectively, at the point A1.  The number of events before
the sideband subtraction only differs by 5\% between HERWIG and
PYTHIA, while the difference increases to be more than 20\% after the
subtraction.  This indicates that the resolution of the $W$ boson mass
is somewhat better for the HERWIG sample.  Indeed, the major
difference of the two generators are in the fragmentation
scheme. PYTHIA is based on the string model, while HERWIG is based on
the QCD model.  Currently we are only able to say that understanding
the nature of fragmentation is essential for the interpretation of
$\nfit$.  We compare the HERWIG and PYTHIA samples in Appendix.

\begin{figure}
\centerline{\psfig{file=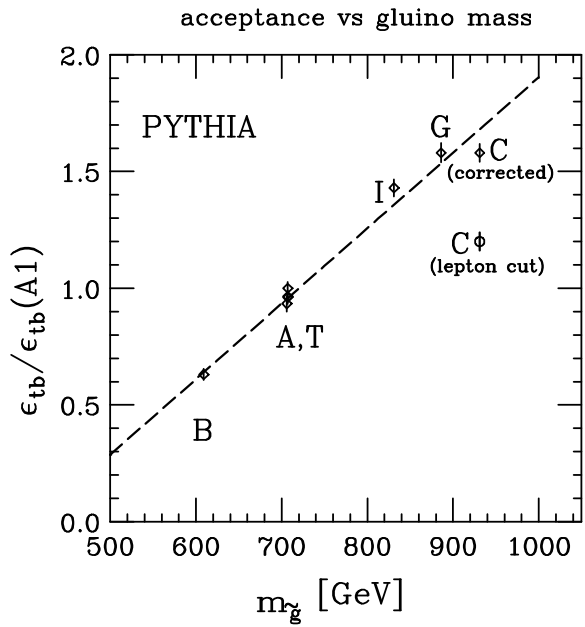,width=3in}} 
\caption{Reconstruction efficiencies
of the model points in Table~\ref{masscross} relative to that 
of the point A1 as a
function of $m_{\tilde{g}}$. Point C is 
plotted twice with/without the loose lepton cut.}
\label{acceptance}
\end{figure}

Let us discuss other effects that might change the reconstruction
efficiency.  In Fig.~\ref{acceptance} we plot the reconstruction
efficiency $\epsilon_{tb}$ as a function of $m_{\tilde{g}}$ by
normalizing the efficiency to that for the point A1.  Here $\nprod$ is
calculated using Eq.~(\ref{nprod}), the generator information on
$N(\tilde{g}\tilde{g})$, $N(\tilde{q}^{(*)} \tilde{g})$, and the Monte
Carlo inputs of the gluino branching ratios.  The length of the bar in
the plot shows twice of the statistical error of $\nedge$,
$\delta(\epsilon_{tb}/{\epsilon}_{tb}(A1))= 2
({\epsilon}_{tb}/{\epsilon}_{tb}({\rm A1}))/\sqrt{\nedge}$.  We don't
discuss the points E1 and E2 here because events with two bottom
quarks are not the dominant signature of squark and gluino production
at these points.

The efficiencies are very close to one another among the points A1,
A2, T1, and T1.  Although the stop mass is very different among the
points, the mass has little influence on the efficiency. Note that
these points have the same gluino masses, and almost the same
parameters for the chargino and neutralino sector.  For the other
points, the efficiency can also be expressed as a linear function of
$m_{\tilde{g}}$ except the sample point C.  The gluino mass dependence
arises from the pre-selection cut $m_{\rm eff}>1000$~GeV.

The reconstruction efficiency at the point C 
(a circle in Fig.~\ref{acceptance}) is lower 
by 20\% from the linear fit (the dashed line in Fig.~\ref{acceptance}).
This is due to the loose lepton cut which reduces the $t\bar{t}$
background.  As the chargino decay branching ratio into leptons is
high (about 47\%) at the point C, the lepton cut kills a significant
fraction of $tb$ events.  The effect is estimated by comparing
$\nedge$ without any lepton cut and $\nedge$ with the loose lepton
cut, at the points C and A1.  The efficiency at the point C without
any lepton cut is plotted in Fig.~\ref{acceptance}, and is close to
the linear fit of the other points.

The gluino mass dependence of the reconstruction efficiency would be
easily corrected by estimating the gluino mass from the effective mass
distribution.  The uncertainty from the chargino decay branching ratio
may be corrected by studying $tb$ final states with leptons as well.
Therefore the parameter dependence of the efficiency may be corrected
by analyzing the Monte Carlo and real data.  Efforts to determine the
MSSM parameters are of course very important to this purpose.

\subsection{Branching ratios}
\label{secbr}

In the previous subsection, we find that the reconstruction
efficiencies significantly differ between the PYTHIA and HERWIG
samples. However, as we can see in Table~\ref{fittable}, the generator
dependence mostly cancels in the ratio $\nfit/\nall$. Therefore the
ratio may play a role to determine the fundamental parameters. For
example, when the contributions from the stop or sbottom pair
productions are negligible, $\nfit/\nall$ is determined by the branching
ratios to the third generation squarks. However, there are several
aspects one must consider.

$\nall$ consists of contributions from various decay modes, even if
the contributions from the stop or sbottom pair productions are
negligible:
\begin{itemize}

\item 
The $t\bar{t}$ final state from the decay chains (II) contributes to
$\nall$.  The branching ratio is typically 1/3 of that of the $tb$
final state for $\mu>M_2>M_1$. The $t\bar{t}$ final state is
reconstructed as $tb$ with a high probability, although the $m_{tb}$
distribution does not have an edge structure.  

\item
We expect that events having no $t \to bW$ decay would be eliminated
by the sideband subtraction, and should not contribute to $\nall$.
However, some events actually remain, because of $W$ and $Z^0$ bosons
from the decay of charginos and neutralinos, as well as accidental $W$
and $Z^0$ bosons from other cascade decay chains.  An example where
$\tilde{\chi}^0_2$ decays dominantly into the $Z^0$ boson will be
discussed in the next subsection.

In addition, the sideband subtraction is not necessarily perfect, as
will be discussed in Appendix. For example, the efficiency of the mode
(I)$_2$ is roughly a half of that of the mode (III)$_1$ after the
sideband subtraction at the point A1 (see Appendix).
\item

The cuts to reduce background events may induce decay mode dependence
of the efficiency.  For example the lepton cut could efficiently
reduce the edge mode (III) if $\Br(\chapm\rightarrow l X )$ is large
as we have seen in Fig.~\ref{acceptance} for point C.

\end{itemize}

The interpretation of the ratio $\nfit/\nall$ becomes more
complicated for the following two cases:
\begin{itemize}
\item
If $m_{\tilde{t}(\tilde{b})}<m_{\tilde{g}}<m_{\tilde{q}}$, squarks
mostly decay into the gluino.  As discussed previously, the
squark/gluino production and decay typically produce events with four
$b$-jets.  They are identified as a two $b$-jet event when two $b$-jets
are tagged and the other two are mis-tagged.  These events decrease
the fraction from the decay chain (III) and (IV) in the two $b$-jet
events.  Assuming a tagging efficiency of 60\% for a single $b$-jet,
35\% of the four bottom quark events are identified as two $b$-jet events,
and only 1/3 of them contain a $b$-jet pair originated from a single
gluino decay. The wrong $b$-jet pair combinations significantly dilute
the edge mode. Study of events with three $b$-jets and four $b$-jets is
necessary for this case.
\item   
The $tb$ signal from the stop pair production could be significant if
the stop is much lighter than the gluino. For example, only 4.7\% of
the SUSY events comes from the stop pair production at the point A1,
while it amounts to be about 17\% at the point T1.
\end{itemize}

In the MSUGRA model, both the right-handed stop and the left-handed
sbottom is lighter than gluino in a broad parameter space. 
As a result, the decay modes which involve $W$ bosons (modes (II), (III) and 
(IV)) dominate over the gluino decays to $bbX$ in the region. 
Because the events with $W$ bosons remain after the $W$ sideband subtraction, 
the reconstruction efficiency $\epsilon_{tb}$ is expected to be 
similar for those decay modes. 
Thus, if the contributions from the stop
or sbottom pair productions are negligible, the numbers of events with
two bottom quarks are given approximately as
\begin{eqnarray}
\nfit &\sim&  {\epsilon_{tb}}\Bredge \times 
\cr
&& \hskip -20pt \left[2 N(\tilde{g}\tilde{g}) 
\left(1-\Br(\tilde{g}\rightarrow bbX)\right)+ N(\tilde{g}\tilde{q})
+ N(\tilde{g}\tilde{q}^*)\right],
\cr
\nall &\sim& {\epsilon_{tb}}\Br(\tilde{g}\rightarrow bbX)\times \cr
&& \hskip -20pt \left[
2N(\tilde{g}\tilde{g}) 
\left(1-\Br(\tilde{g}\rightarrow bbX)\right)
+ N(\tilde{g}\tilde{q})+ N(\tilde{g}\tilde{q}^*)
\right].\cr&&
\end{eqnarray}

If this simple
formula holds, the ratio $\nfit/\nall$ should provide the ratio of
the branching ratios $\Bredge/\Br(\tilde{g}\rightarrow bbX)$.

\begin{figure}
\centerline{\psfig{file=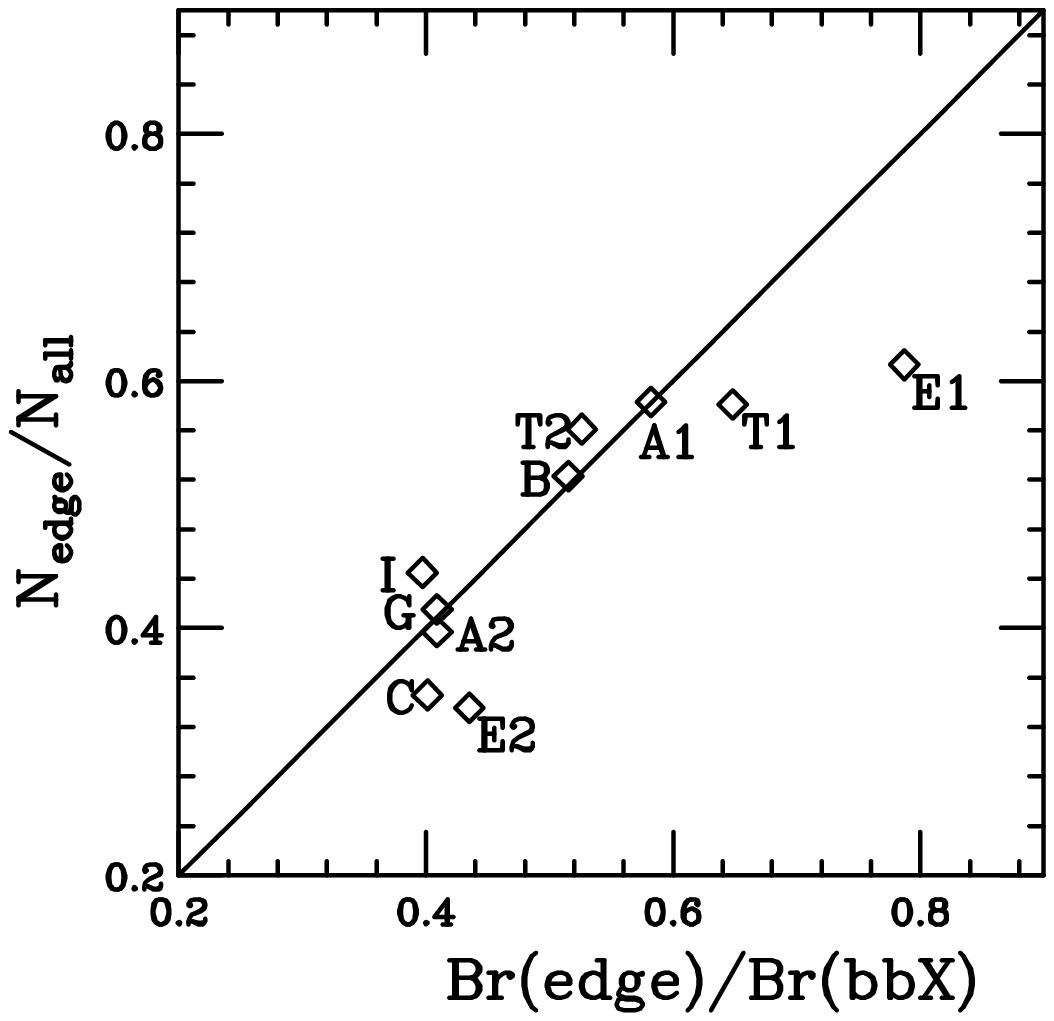,width=3in}}
\caption{Relation between
$\nedge/\nall$ and $\Bredge/\Br(\tilde{g}\rightarrow bbX)$.}
\label{ratio}
\end{figure}

This is illustrated in Fig.~\ref{ratio}, where we plot the ratio
$\nedge/\nall$ as a function of $\Bredge / \Br(\tilde{g}\rightarrow
bbX)$.  Here we plot $\nedge/\nall$ instead of $\nfit/\nall$, because
we have already seen $\nfit\sim \nedge$, and the statistical
fluctuation of $\nedge$ is small ($\sim$ 2\%) with a help of the
generator information.  Some points in the plots are away from the
line $\nedge/\nall= \Bredge /\Br(\tilde{g}\rightarrow bbX)$: The point
C is off because the chargino has large branching ratios into leptons
as discussed earlier.  At the point T1, the
$\tilde{t}_1\tilde{t}^{*}_1$ productions contributes to $\nall$. The
points E1 and E2 are significantly off because the first and the
second generation squarks dominantly decay into the gluino.

The systematics due to the chargino and neutralino decay patterns are
difficult to estimate from the LHC data only.  By combining the data
from a future $e^+ e^-$ linear collider (LC) of $\sqrt{s}=500 \sim
1000$~GeV such systematics would be significantly reduced.  Note that
SUSY points with $m_{\chap}\lsim 300$ GeV are studied in this paper,
which is within the reach of a TeV scale LC.

If events with four $b$-jets dominate the SUSY events at the LHC,
study of the events with four or three $b$-jets would be important,
which is out of the scope of this paper.  If $m_{\stp}\ll
m_{\tilde{g}}$, it is important to constrain the stop mass in order to
estimate the contribution from $\tilde{t}_1\tilde{t}_1^*$, as the
cross section increases rapidly as $m_{\stp}$ decreases.  The weighted
end point $\mtbw$ would provide the information on $m_{\stp}$.

\subsection{Snowmass points}
\label{Ssnowmass}

In this subsection we check if we would reconstruct fake end points by
any chance when the mode of our interest is insignificant.  To this
purpose, we generate a set of Monte Carlo data for the Snowmass
points~\cite{snowmass}.  The Snowmass points and slopes (SPS) are a
set of benchmark points and parameter lines in the MSSM parameter
space corresponding to different scenarios in the search for SUSY at
the present and future experiments.  Some of those points correspond
to the case where decay modes (III) or (IV) are not dominant, and we
should not expect to see the edge structure in the $m_{tb}$
distribution.

We list the fit results at SPS~1-2, and 4-6 in Table~\ref{snowmass}.
Here SPS~3 is not listed as it is the same as the point C.

The $\mtbfit$ is lower by 20~GeV than the $\mtbw$ at SPS~1 and SPS~6,
similar to the points discussed in Subsection~\ref{kinematical} (see
Table~\ref{fittable}).

\begin{figure}
\centerline{
\psfig{file=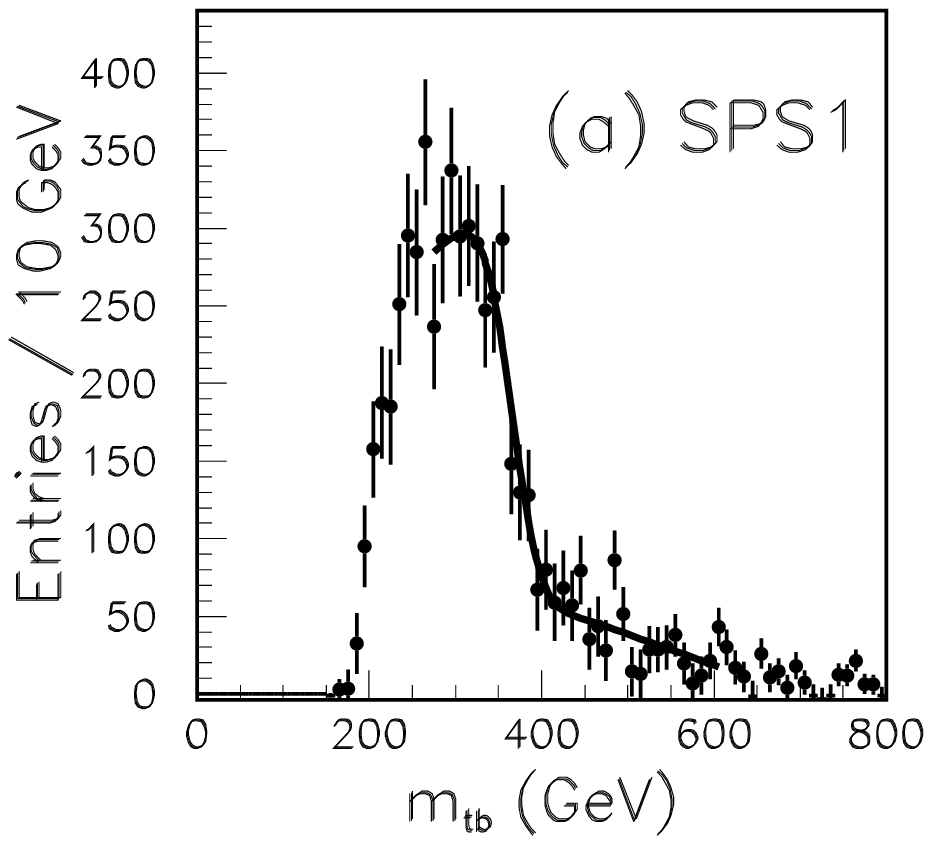,width=3in} 
}
\vspace{-1cm}
\centerline{
\psfig{file=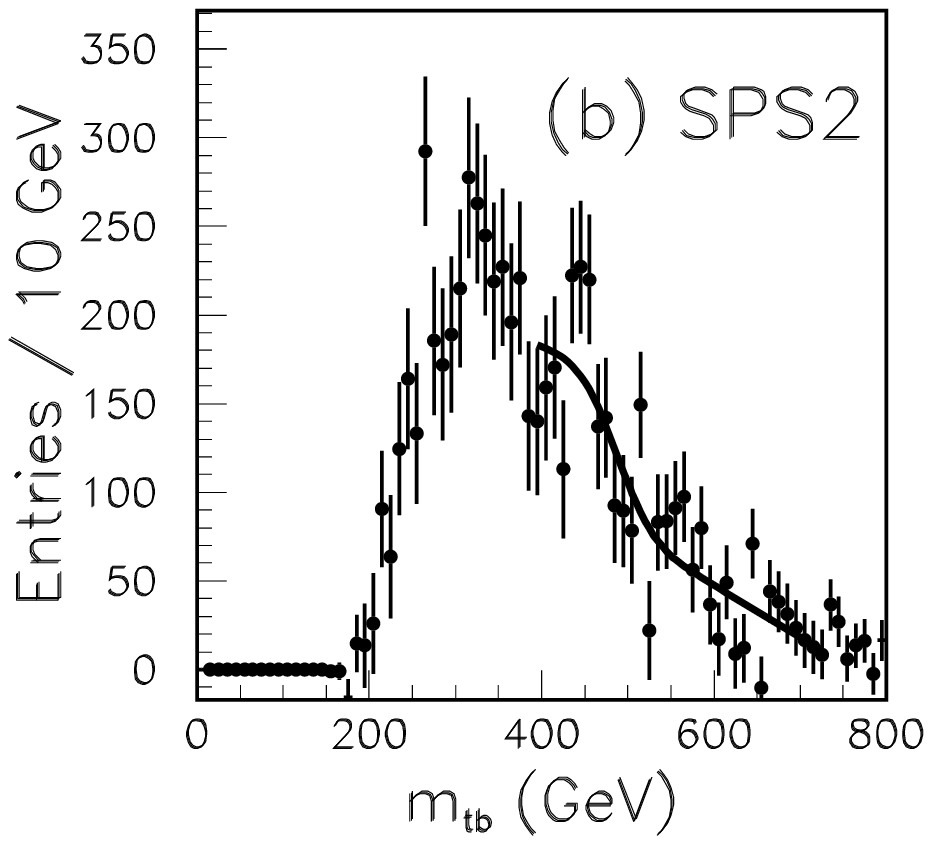,width=3in} 
}
\caption{The $m_{tb}$ distributions at
(a) SPS~1 and (b) SPS~2.
The fit curves are also shown.}
\label{SPS12}
\end{figure}

SPS~2 is so-called the ``focus point'', where the scalar mass at the
GUT scale $m_0$ is larger than the GUT scale gaugino mass $M_0$.  The
gluino two body decays into a squark and a quark are not kinematically
open, therefore the dominant decay modes of the gluino are the three
body decays; $\Br(\tilde{g}\rightarrow tb\chaspm ) =25\% $,
$\Br(\tilde{g}\rightarrow tb\chapm ) =20\% $, and
$\Br(\tilde{g}\rightarrow tt{\tilde{\chi}_i}^{0'}s)=20\%$.  The
$m_{tb}$ distribution should not have any edge structure because it
dominantly consists of the gluino three body decays.  The
reconstructed edge has only $4\sigma$ in height, and is statistically
insignificant.  The $\chi^2$ value of the fit is also rather large,
$\chi^2=$44.7 for 30 degrees of freedom
\footnote{However, 
$\Delta \chi^2$ is sometimes large even when the decay mode (III)
dominates the distribution.  For example $\Delta\chi^2=41.4$ for the
point A1.}.  The fitted end point $\mtbw= 484.9\pm 24.9$~GeV may be
related to the mass difference between the gluino and the heavier
chargino, $m_{\tilde{g}}=778.6$~GeV and
$m_{\tilde{\chi}^{\pm}_2}=296.3$~GeV.  The $m_{tb}$ distribution at
SPS~2 is compared with that at SPS~1 in Fig.~\ref{SPS12}.

At SPS~5, the stop is so light that the decay $\tilde{t}_1 \rightarrow
b\tilde{\chi}_1^{\pm}$ is kinematically closed.  The gluino decay
branching ratio into $\tilde{b}_1$ is large (9\%), but the
$\tilde{b}_1$ subsequently decays mostly into $W\tilde{t}_1$(80\%).
We thus do not find any significant edge.

\begin{figure}
\centerline{
\psfig{file=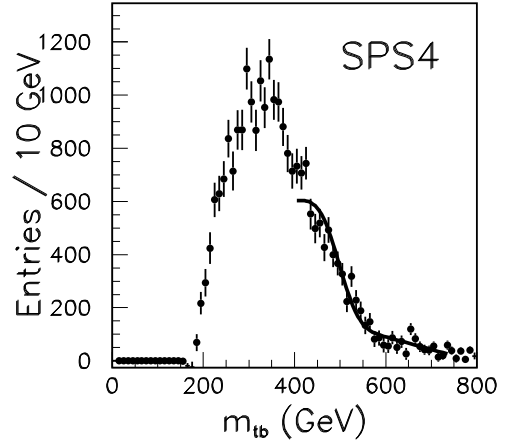,width=3in}
}
\caption{The $m_{tb}$ distribution at SPS~4.
The fit curve is also shown.}
\label{SPS4}
\end{figure}

SPS~4 is an example that a straightforward application of our
reconstruction technique fails to find $\mtbw$.  The distribution is
shown in Fig.~\ref{SPS4}.  The sbottom is light because of large
$\tan\beta$ , and the dominant gluino decay mode is
$\tilde{g}\rightarrow b\tilde{b}_1$ (78\%).  The sbottom further
decays into chargino (36\%) or the second lightest neutralino (30\%).
Decays of the chargino and the second lightest neutralino are
dominated by $\tilde{\chi}^{\pm}_1\rightarrow W\tilde{\chi}^0_1$
(100\%) and $\tilde{\chi}^0_2\rightarrow Z^0\tilde{\chi}^0_1$ (98\%),
respectively.  The dominance of the decay into $Z^0$ or $W$ is common
when the decay into the Higgs boson or sleptons are kinematically
forbidden.  Some of the decays $\tilde{g}\rightarrow b
\tilde{b}\rightarrow b b \tilde{\chi}^0_2$ are reconstructed as $tb$
events together with additional gauge bosons, resulting in the
$m_{tb}$ distribution having a quasi-edge structure.  The edge is
reconstructed by our fit (see Table~\ref{snowmass}).  Note that the
end point of the $m_{bb}$ distribution of the decay is 403~GeV, while
the reconstructed $m_{tb} $ distribution of the events with the decay
(I)$_{2}$ has a peak around 500~GeV and the distribution ends around
580~GeV, close to the kinematical limit
$m_{\tilde{g}}-m_{\tilde{\chi}}\sim 600$~GeV.  The edge structure is
identified as that of the mode (I)$_{2}$ with a tagged $Z^0\rightarrow
ll$.  In the $m_{tb}$ distribution, the second edge structure from the
decay mode (III) or (IV) is weakly seen in the distribution. Assuming
that two edges exist, we find the second end point $\mtbfit=425.4\pm
15.2$~GeV and $h=349.4\pm 74.4$/(10~GeV).

\begin{table}[ht]
\begin{tabular}{|c||rrr|}
\hline
     &$\mtbw$[GeV] & $M^{\rm fit}_{tb}$[GeV] & $h$/(10~GeV)  \cr
\hline
SPS~1 &380.8& 363.9 $\pm$ 4.8  & 267.3 $\pm$ 20.8 \cr
SPS~2 & N/A         & 484.9 $\pm$ 24.9 &  92.2 $\pm$ 23.5 \cr
SPS~4 & 431.0       & 495.0 $\pm$ 6.2  & 515.7 $\pm$ 33.0 \cr
SPS~5 &416.2        & 442.6 $\pm$ 42.8 &  56.6 $\pm$ 69.6 \cr
SPS~6 &442.3        & 416.2 $\pm$ 6.2  & 326.4 $\pm$ 24.7 \cr
\hline
\end{tabular}
\caption{Fit results  for the Snowmass points. 
Data sample at SPS~2 corresponds to $2\times 10^6$ events, 
while other samples correspond to   $3\times 10^6$ events.}
\label{snowmass}
\end{table}
 
\section{Events with Additional Tags} 

At the points given in Table~\ref{masscross}, the lighter chargino and
neutralinos (inos) $\tilde{\chi}^{\pm}_1$, $\tilde{\chi}^{0}_1$ and
$\tilde{\chi}^{0}_2$ are gaugino-like, while the heavier inos
$\tilde{\chi}^{h}\equiv$ $\tilde{\chi}^{\pm}_2$, $\tilde{\chi}^{0}_3$
and $\tilde{\chi}^{0}_4$ are higgsino-like.  Unlike the first and
second generation squarks, the third generation squarks can decay into
the heavier inos.  This is because the third generation squarks couple
to the higgsinos by the top (or bottom) Yukawa coupling.  The
identification of the cascade decay chain $\glu\rightarrow
\tilde{t}/\tilde{b} \rightarrow$ $\tilde{\chi}^h$ is a probe for the
supersymmetric version of Yukawa interactions.  In addition, the decay
distribution may be sensitive to the mass difference
$\mglu-m_{\tilde{\chi}^h}\sim M_{3}-\mu$.

In this section we show an example to select such events by requiring
additional leptons in the final state.  We study the point C, where
the stop and sbottom decay into $\tilde{\chi}^h$ with large branching
fractions.  At this point the gluino decay branching ratios are
Br$(\glu\rightarrow\stp)=15.1$\% and Br$(\glu\rightarrow\sbt)=14$\%.
The squarks further decay into the higgsino-like inos with branching
ratios of $\Br(\sbt\rightarrow \tilde{\chi}^h)=24$\% and
$\Br(\stp\rightarrow \tilde{\chi}^h)=13$\%, respectively.  The stop
and sbottom decay branching ratios are comparable to those into
$\chapm$.

Some of the decay branching ratios of the charginos and neutralinos
are listed in Table~\ref{pointcbr}.  The higgsino-like inos have large
branching ratios into the $Z^0$ boson.  The heavier inos also have
non-negligible branching ratios into triple leptons, because they may
also decay into $\tilde{\chi}^0_2$ which further decays into leptons.

\begin{figure}
\centerline{\psfig{file=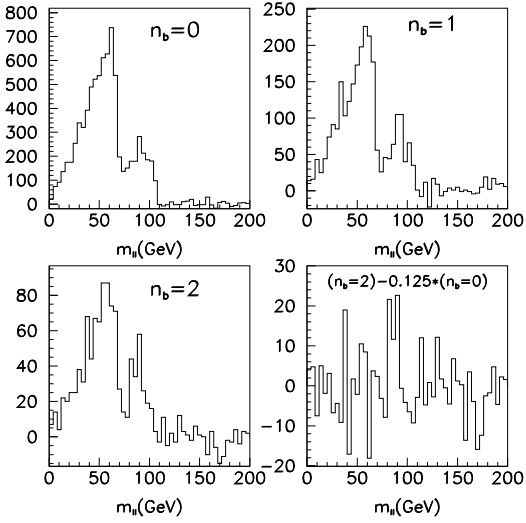,width=3in}} 
\caption{Distributions of $m_{e^+e^-}+m_{\mu^+\mu^-}-m_{e^+\mu^-}
-m_{e^-\mu^+}$ with $n_{b}=0$ (top-left), 1 (top-right),
2 (bottom-left). The bottom-right plot shows the distribution for
$n_{b}=2$ after subtracting the scaled distribution for $n_b=0$  by 
a factor of 0.125.}
\label{ll}
\end{figure}

The signature of the heavier inos from stop or sbottom is an excess of
the $Z^0$ boson or three leptons in events with two $b$-jets.  In
Fig.~\ref{ll} we plot the invariant mass of the same flavor opposite
sign leptons ($m_{ll}$) for the events with $n_b=0, 1 $ and 2, where
$n_b$ is the number of $b$-jets in an event.  In the plots, accidental
lepton pair distribution estimated with events with different-flavor
opposite sign leptons are subtracted.  All plots show a common
structure that the invariant mass distribution increases toward the
edges around 70~GeV and 105~GeV, which correspond to the decay chains
of $\neus\rightarrow l\tilde{l}_{L} \rightarrow ll\neu $ and
$\neus\rightarrow l\tilde{l}_{R} \rightarrow ll\neu$, respectively.
The $Z^0$ peak is also seen in the plots.  As $n_b$ increases, the
$Z^0$ peak height becomes more significant relative to the edge
structure.  The excess of $Z^0\rightarrow ll$ in the $n_b=2$ sample
indicates the $Z^0$ boson originating primarily from stop or sbottom
Qualitatively, the edges in the $m_{ll}$ distribution for the $n_b=2$
sample are completely subtracted by the distribution for the $n_b=0$
sample scaled by a factor of 0.125, and the $Z^0$ peak remains (the
bottom-right plot of Fig.~\ref{ll}).

\begin{table}
\begin{tabular}{| l|c|l|c|} 
\hline
decay mode & Br in \% & decay mode & Br in \%\cr
\hline
$\chaspm\rightarrow W\neus $ & 29&
$\neuth\rightarrow W\tilde{\chi}^{\pm} $ & 59\cr
$\chaspm\rightarrow Z^0\chapm $ & 26 & 
$\neuth\rightarrow Z^{0}\neu $ & 1.7\cr
$\chaspm\rightarrow l\tilde{\nu} $ &3.0 &
$\neuth\rightarrow Z^0\neus $ &25\cr
$\chaspm\rightarrow \nu\tilde{l} $ & 6.8 &&\cr 
\hline
$\neufr\rightarrow W\tilde{\chi}^{\pm}_1 $ & 59&
$\neus\rightarrow l\tilde{l}_L $ & 9.8 \cr
$\neufr\rightarrow Z^0\neu $ &1.7&
$\neus\rightarrow l\tilde{l}_R $ & 3.0\cr
$\neufr\rightarrow Z^0\neus $ &1.3&
$\chapm\rightarrow W\neu $ & 11\cr
$\neufr\rightarrow l\tilde{l}_L $ & 3.0&
$\chapm\rightarrow l\tilde{\nu}_L $ & 37\cr
$\neufr\rightarrow l\tilde{l}_R $ & 0.9&
$\chapm\rightarrow \nu \tilde{\l}_L $ & 8.9\cr
\hline
\end{tabular}
\caption{Some of the chargino and neutralino branching ratios in \% 
at the point C. }
\label{pointcbr}
\end{table}

\begin{figure}
\centerline{\psfig{file=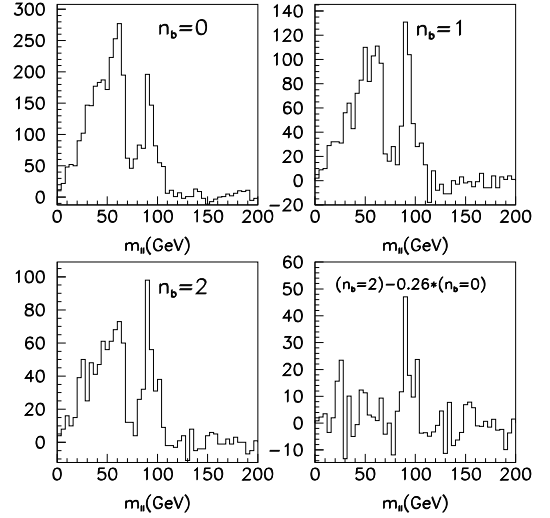,width=3in}} 
\caption{ Same as Fig.~\ref{ll}, but for events with three leptons. 
}
\label{lll}
\end{figure}

The same analysis can be performed by using events with three
leptons. In Fig.~\ref{lll} we plot sum of invariant mass
distributions; $m_{e^+e^-}+m_{\mu^+\mu^-}- m_{e^{+} \mu^{-}} -m_{e^{-}
\mu^{+}} $ for all possible combinations of opposite sign lepton
pairs.  In this mode the $Z^0$ peak is more significant.  The ratio of
the edge heights in the $m_{ll}$ distribution for the $n_b=2$ sample
to that of the $n_b=0$ sample is 0.25.  This means that the events
originating from the decay
$\tilde{b}/\tilde{t}\rightarrow\tilde{\chi}^h$ dominate the three
lepton events, and the events with $n_b=0$ arise due to misidentified
$b$-jets.

Note that these plots at the point C are made for a total of $3\times
10^6$ SUSY events, which corresponds to an integrated luminosity of
600 fb$^{-1}$.  The statistics for 100 fb$^{-1}$ is not significant.
However, this procedure should be useful for lighter mass spectrum
where the decay of the third generation squarks to $\tilde{\chi}^h$ is
open.

\begin{figure}
\centerline{\psfig{file=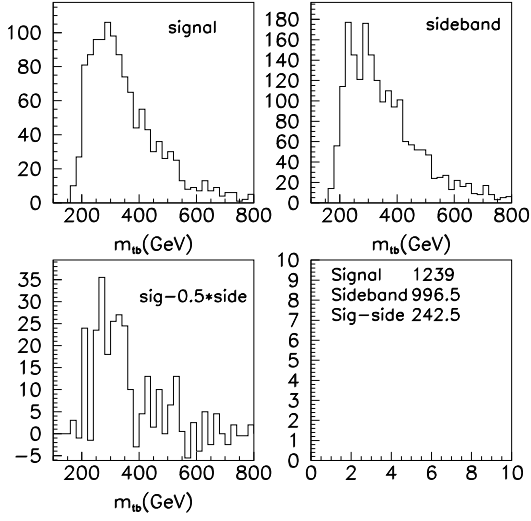,width=3in}}
\caption{The $m_{tb}$ distributions for events
with tagged three leptons. Top-left: 
signal distribution. Top-right:  background estimated 
by the sideband events. Bottom-left: distribution after the sideband 
subtractions.}
\label{tb3l}
\end{figure}

After establishing the contribution of the decay
$\tilde{t}/\tilde{b}\rightarrow\tilde{\chi}^h$ in the tagged $bb3l$
and $bbZ^0$ events, we may reconstruct the $tb$ system in the sample.
The $tb$ reconstruction is similar to that described in Section~II,
except that the distributions are made without the cuts for the
effective mass and the lepton, because the standard model background
is expected to be small now.  In Fig.~\ref{tb3l} the $m_{tb}$
distribution with tagged three leptons is shown. The number of events
after the sideband subtraction is 242.5.  The $m_{tb}$ distribution is
concentrated just below the expected end point which is close to
$\mglu-m_{\chaspm}\sim 400$~GeV, although the statistics is low. The
distribution of tagged $Z^0\rightarrow ll$ is similar and the number
of events after the sideband subtraction is 142.

\section{Stop and Sbottom Properties in the MSUGRA Model}

In this section we interpret our study of the third generation squarks
at the LHC in a framework of the MSUGRA model.  The LHC may determine
the MSUGRA parameter $m_0$, $M_0$, $A_0$, and $\tan\beta$ through the
measurement of various distributions. The decay chains $\tilde{q}
\rightarrow q \tilde{\chi}^0_2$ followed by $\tilde{\chi}^0_2
\rightarrow$ $h \tilde{\chi}^0_1$,$\ Z^0 \tilde{\chi}^0_1$,
$l\tilde{l}$, or $ll\tilde{\chi}^0_1$ are especially useful if the
kinematical end points of the distributions are measured. The
effective mass measurement of the inclusive SUSY signal is also useful
to determine the absolute SUSY scale.  Among these parameters, the
$A_0$ parameter is only weakly constrained.  This is because the
left-right mixings of sbottom and stau are almost fixed by
$\mu\tan\beta$, and $A_t$ is insensitive to $A_0$ as will be described
below.  One of the possible ways to determine $A_0$ is to measure the
mass difference between squarks or sleptons with different flavors,
which is caused by the renormalization group equation (RGE) running
between the GUT scale and the weak scale.

The squark-mass matrix is given as follows,
\begin{eqnarray}
-{\cal L}_{\rm mass} &=&
\left(
\begin{array}{cc}
\tilde{f}_L^\star &
\tilde{f}_R^\star
\end{array}
\right)
\left(
\begin{array}{cc}
m_{LL}^2&m_{LR}^2\\
m_{LL}^{2} &m_{RR}^2
\end{array}
\right)
\left(
\begin{array}{c}
\tilde{f}_L\\
\tilde{f}_R
\end{array}
\right),
\end{eqnarray}
where
\begin{eqnarray}
m_{LL}^2 &=&
m_{\tilde{t}_L}^2+m_t^2+m_Z^2 \cos 2\beta \,(\frac12-\frac23 \sin^2\theta_W),
\nonumber\\
m_{RR}^2 &=&
m_{\tilde{t}_R}^2+m_t^2+ \frac23 m_Z^2 \cos 2\beta  \sin^2\theta_W,
\nonumber\\
m_{LR}^2 &=&
m_t\,(A_t-\mu \cot\beta)
\end{eqnarray}
for stops, and 
\begin{eqnarray}
m_{LL}^2 &=&
m_{\tilde{t}_L}^2+m_b^2+m_Z^2 \cos 2\beta \,(-\frac12+\frac13 \sin^2\theta_W),
\nonumber\\
m_{RR}^2 &=&
m_{\tilde{b}_R}^2+m_b^2- \frac13 m_Z^2 \cos 2\beta  \sin^2\theta_W,
\nonumber\\
m_{LR}^2 &=&
m_b \,(A_b-\mu \tan\beta)
\end{eqnarray}
for sbottoms.  

In the MSUGRA, the SUSY-breaking parameters at the weak scale in these
mass matrices are evaluated by the RGEs
with the universal GUT scale boundary conditions. 
When the bottom-Yukawa coupling is negligible, it is
convenient to present the SUSY-breaking parameters by the
infrared-fixed point value for the top-Yukawa coupling constant
$\bar{y}_f$\cite{Bardeen:1993rv},
\begin{eqnarray}
\bar{y}_f
&=&
\frac{F'(t)}{6 F(t)}.
\end{eqnarray}
Here,
\begin{eqnarray}
F(t)=\int^t_0 d t'
\left(\frac{\alpha_3(t)}{\alpha_{GUT}}\right)^{\frac{16}{9}}
\left(\frac{\alpha_2(t)}{\alpha_{GUT}}\right)^{-3}
\left(\frac{\alpha_1(t)}{\alpha_{GUT}}\right)^{-\frac{13}{99}}
\end{eqnarray}
where $t=1/(4\pi) \log \mu^2/M_{GUT}^2$ and $\mu$ is the 
renormalization scale. 
$\alpha_{GUT}$ is the
gauge coupling constant at the GUT scale. The SUSY breaking parameters
for the third generation at the low energy scale 
are approximately given using the value as 
\begin{eqnarray}
m_{\tilde{t}_L}^2
&\simeq&
 0.7 M_{\tilde{g}}^2+0.5 m_0^2
-(0.16 A_0^2-0.25 A_0 M_{\tilde{g}})(1-\xi),
\nonumber\\
m_{\tilde{t}_R}^2
&\simeq&
 0.5 M_{\tilde{g}}^2+ m_0^2 (1- \xi) \nonumber\\
&&-(0.33 A_0^2-0.51 A_0 M_{\tilde{g}})(1-\xi) ,
\nonumber\\
m_{\tilde{b}_R}^2
&\simeq&
m_0^2+ 0.8 M_{\tilde{g}}^2,
\nonumber\\
A_t &\simeq& -0.7 M_{\tilde{g}} + A_0 (1-\xi), 
\nonumber\\
A_b &\simeq& -  1.2 M_{\tilde{g}}+ 0.8 A_0
\label{rg}
\end{eqnarray}
up to ${\cal O}((1-\xi)^2)$, 
where 
\begin{equation}
\xi\  (\equiv y_t^2/\bar{y_t}^2)
\ =\left(\frac{m_t}{203 \sin\beta\ ({\rm GeV})}\right)^2.
\end{equation}
Here we use the GUT relation for the gaugino masses, $M_{\tilde{g}}
\simeq 2.8 M_{0}$. For the pole top mass $m_t^{pole}=175$~GeV, 
$\xi$ ranges $0.7\lsim \xi\lsim 0.9$.

From Eq.~(\ref{rg}), the qualitative behavior of the SUSY-breaking
parameters is clear.  The reduction of the $m_{\tilde{t}_L}^2$ and
$m_{\tilde{t}_R}^2$ is more efficient for $A_0 M_{\tilde{g}}<0$,
although the dependence is suppressed by an overall factor of
$1-\xi$. The right handed stop mass $m_{\tilde{t}_R}$ is also
insensitive to $m_0$.  In a moderate $\tan\beta$ region, $\tilde{t}_1$
and $\tilde{b}_1$ is almost right-handed and left-handed,
respectively.

\begin{figure}[ht]
\centerline{
\psfig{file=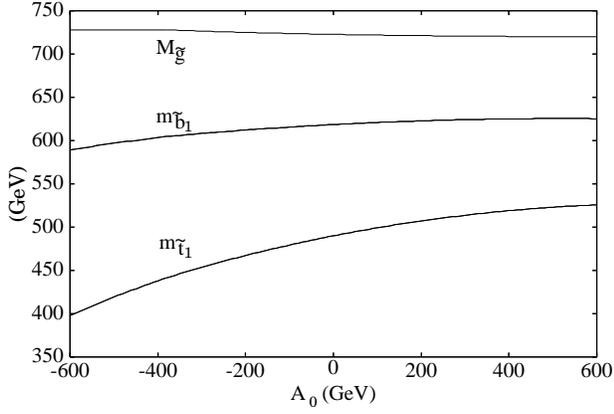,width=3.2in}} 
\caption{Masses for $\tilde{g}$,
${\stp}$, and ${\sbt}$ as functions of $A_0$. Here,
$m_{0} =230$~GeV, $M_{0} =300$~GeV, $\tan\beta=30$, and $\mu>0$. }
\label{msugrafig1}
\end{figure}

In Fig.~\ref{msugrafig1} we show the masses of $\tilde{g}$,
$\tilde{t}_1$, and $\tilde{b}_1$ as functions of $A_0$, where the
other MSUGRA parameters are fixed to be $m_{0} =230$~GeV, $M_{0}
=300$~GeV, $\tan\beta=10$, and $\mu>0$. As expected from the
discussion above, the masses $m_{\tilde{t}_1}$ and $m_{\tilde{b}_1}$
have sensitivity to $A_0$.  The reduction of $m_{\tilde{t}_L}^2$ and
$m_{\tilde{t}_R}^2$ by the top-Yukawa coupling constant is larger when
$A_0 <0$.

\begin{figure}[ht]
\centerline{
\psfig{file=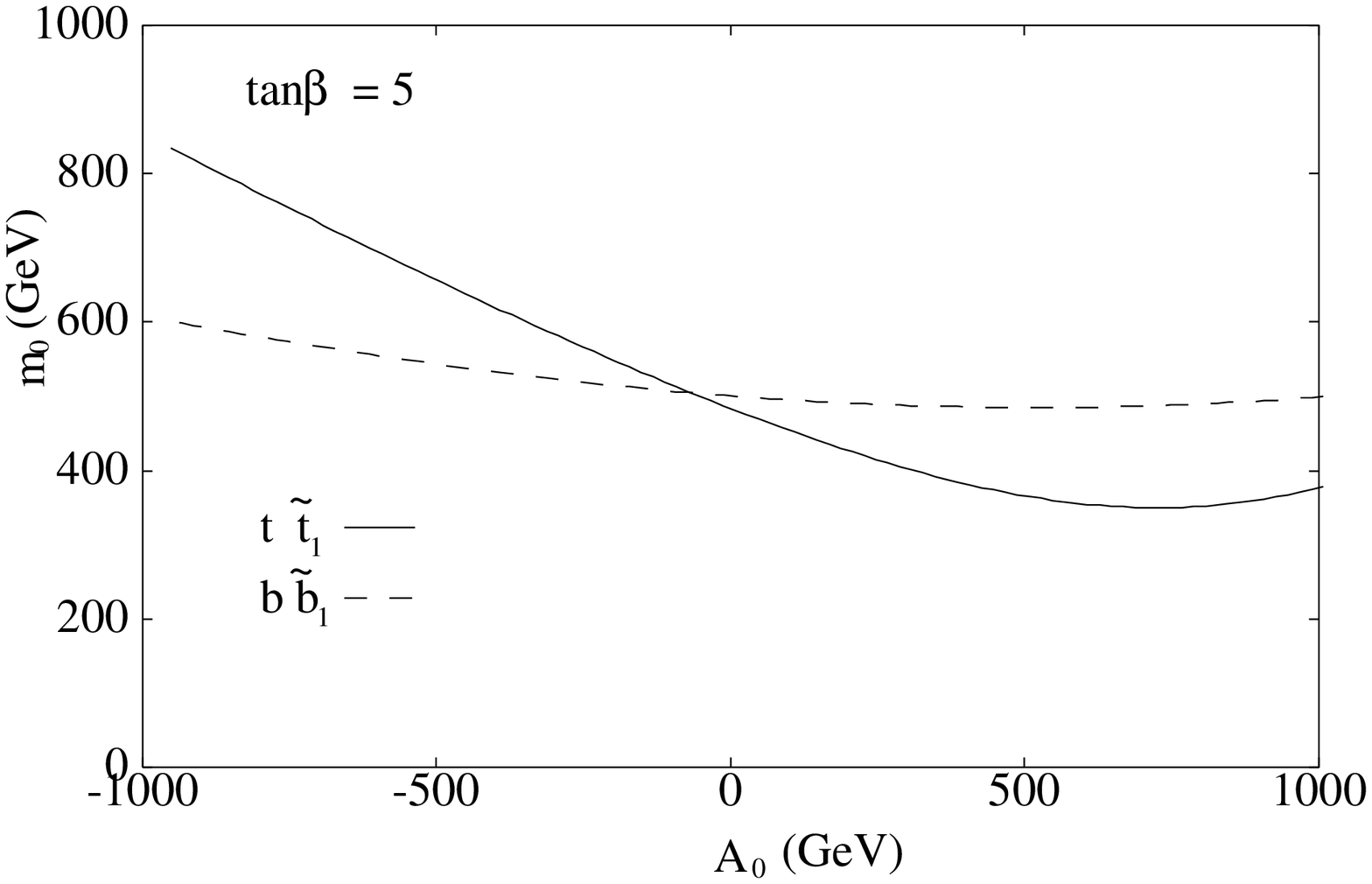,width=3in}} 
\centerline{
\psfig{file=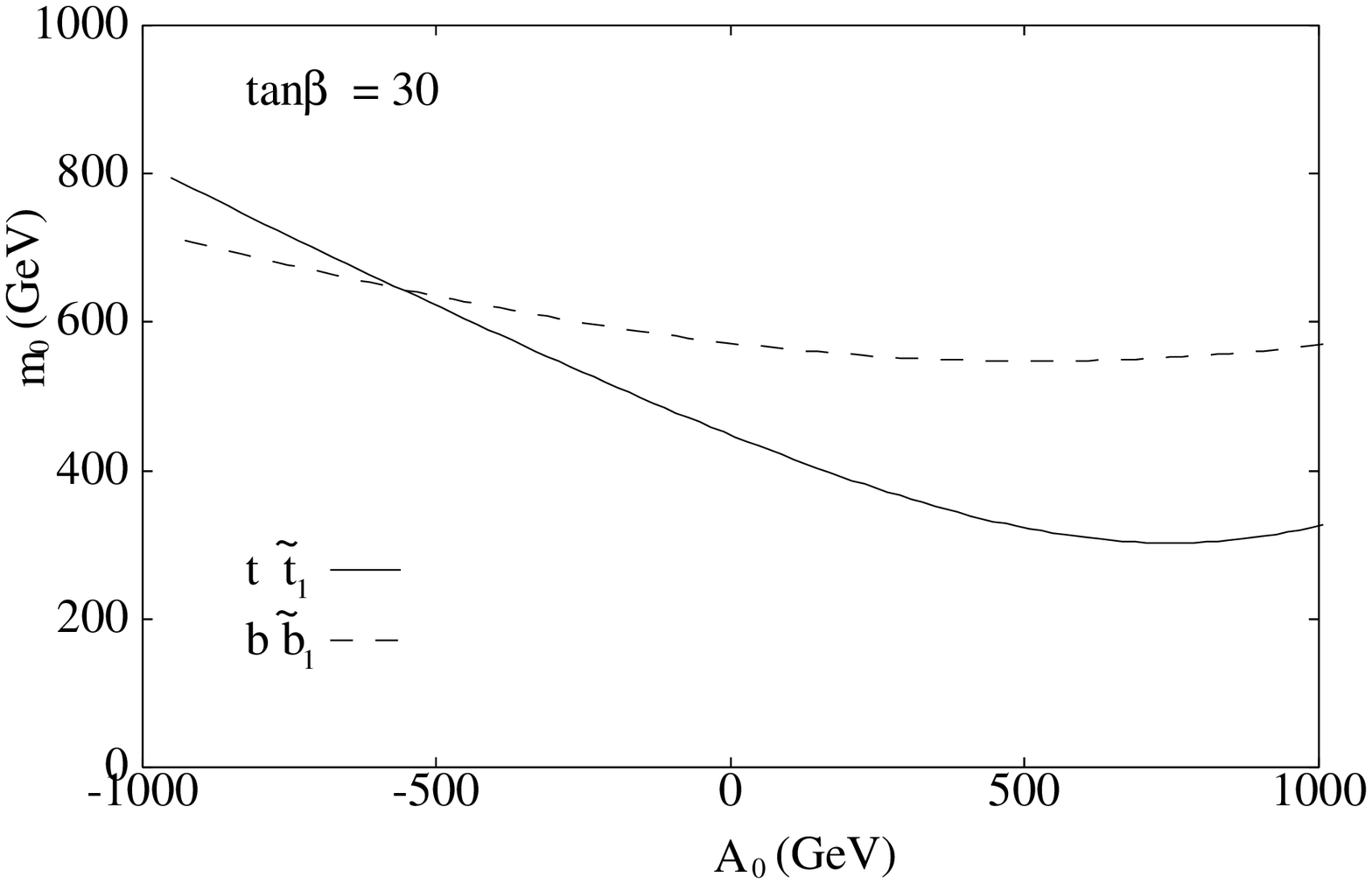,width=3in} 
}
\caption{
Contours where $m_{\tilde{g}}=m_{\stp}+m_{t}$ 
and $m_{\sbt}+m_{b}$ on a plane of
$m_0$ and $A_0$. The gluino decays into $t\stp$ and $b\sbt$
are open below the contours. Here, $m_{\tilde{g}}=707$~GeV, $\tan\beta=$5 and 30, and
$\mu>0$.  }
\label{msugrafig2}
\end{figure}

The stop $\stp$, which is almost $\tilde{t}_R$, is lighter than the
gluino unless $m_0$ is very large.  Similarly, since the sbottom
$\sbt$ is almost $\tilde{b}_L$, the branching ratio to the edge events
in Eq.~(\ref{gluinodecay}), which includes $\chapm$, is large since it
has SU(2) gauge and top-Yukawa interactions.  The study of the edge
events has therefore potentially high feasibility in the MSUGRA.
Fig.~\ref{msugrafig2} shows the regions, where the gluino decays into
$t\stp$ and $b\sbt$ are open, in a plane of $m_0$ and $A_0$.  In this
plot the other parameters are fixed to be $m_{\tilde{g}}=707$~GeV,
$\tan\beta=$5 or 30, and $\mu>0$.

\begin{figure}[ht]
\centerline{
\psfig{file=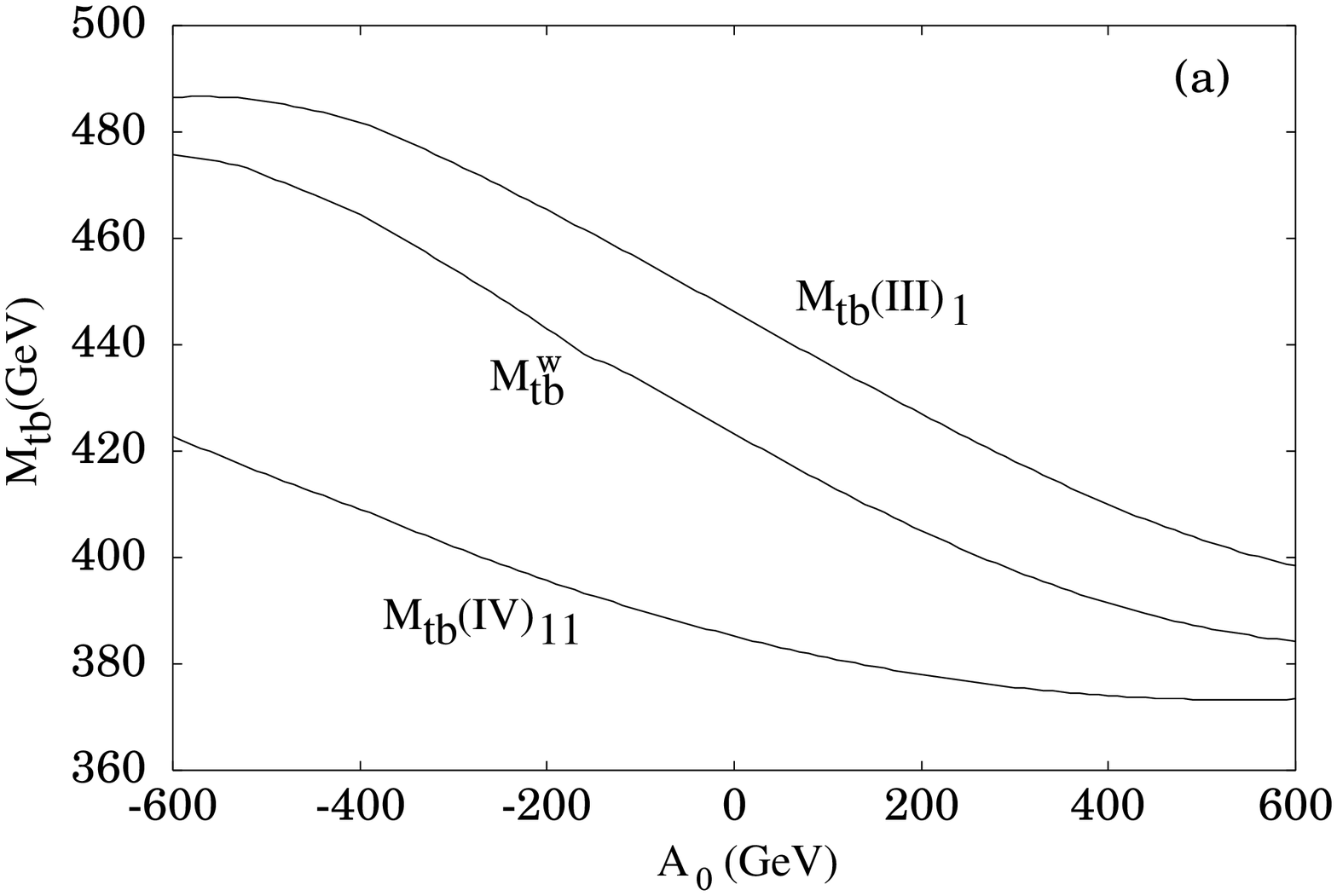,width=3.2in}} 
\centerline{
\psfig{file=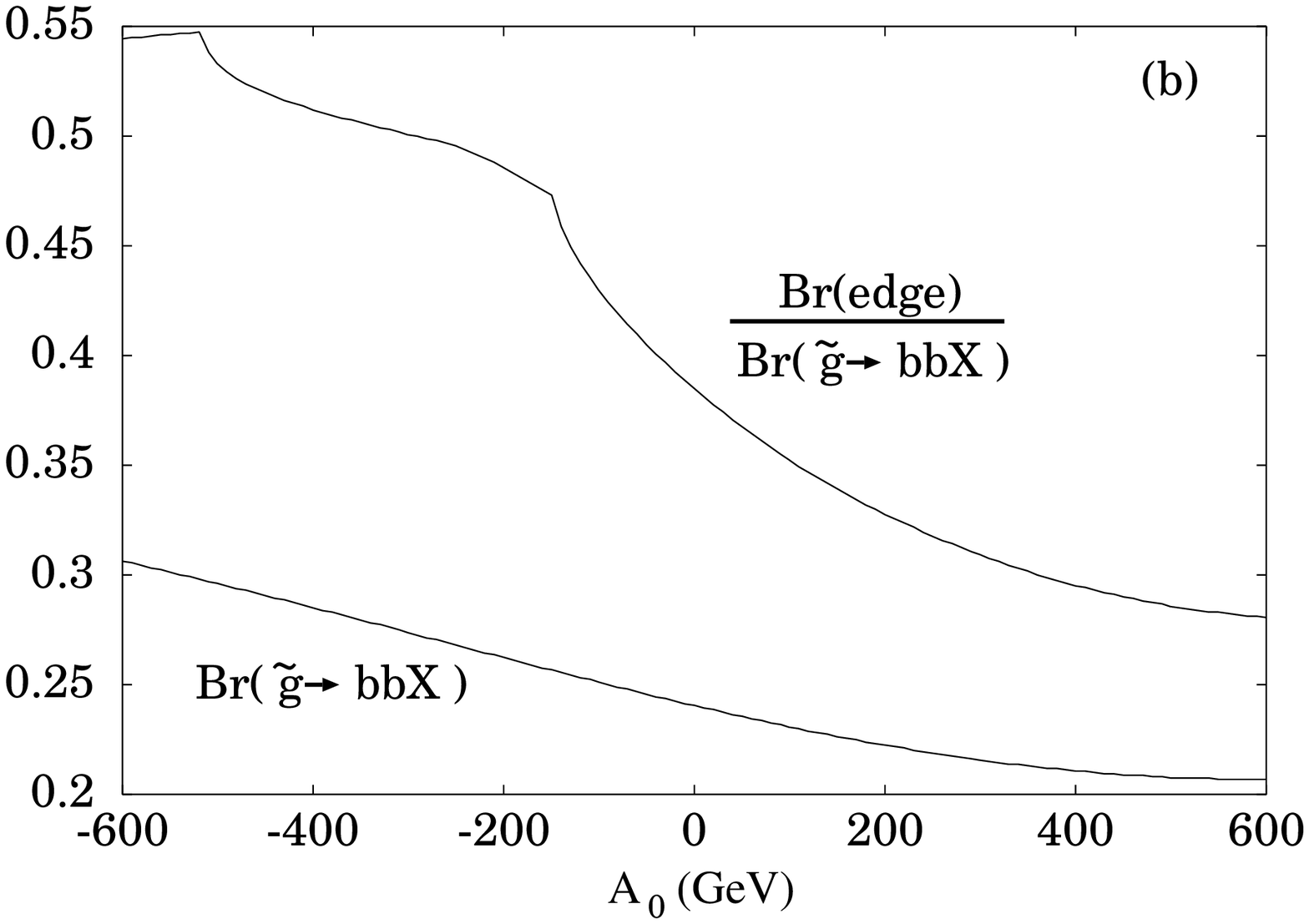,width=3in} 
}
\caption{(a) $M_{tb}^{\rm w}$, $M{\rm (III)}_1$, and  $M{\rm (IV)}_{11}$ 
as functions of $A_0$. (b) $\Br(\tilde{g}\rightarrow bb X)$
and $\Bredge/\Br(\tilde{g}\rightarrow bb X)$ as 
functions of $A_0$. Input parameters are the same as in 
Fig.~\ref{msugrafig1}.}
\label{msugrafig3}
\end{figure}

Since the top quark mass deviates from the fixed-point value, the
$A_0$ dependence on $m_{\tilde{t}_L}^2$ and $m_{\tilde{t}_R}^2$
survives, as shown in Fig.~\ref{msugrafig1}. This may lead to a
measurable dependence on $A_0$ of the end point of the $m_{tb}$ distribution. 
In Fig.~\ref{msugrafig3}(a), 
we show 
$M_{tb}^{\rm w}$, 
$M_{tb}{\rm (III)}_1$, and 
$M_{tb}{\rm (IV)}_{11}$ as functions of $A_0$. 
Since $\sbt$
is heavy, 
$M_{tb}{\rm (IV)}_{11}$ is lower than $M_{tb}{\rm (III)}_{1}$. 
However, Br$(\tilde{g}\rightarrow t\stp )$ is larger than
Br$(\tilde{g}\rightarrow b\sbt )$ due to the phase space, which
makes $M_{tb}^{\rm w}$ closer to $M_{tb}{\rm (III)}_1$. If $M_{tb}^{\rm w}$ is
determined with a precision of a few $\%$ as in  our 
simulation study, 
the $A_0$ parameter can be evaluated with a precision of $\sim 50$~GeV.

In addition to the weighted end point $M_{tb}^{\rm w}$, the edge
height $h$ of the $m_{tb}$ distribution is measurable, and the height
is roughly proportional to the branching ratio of the gluino to the
edge events arising from the decay modes (III) and (IV) in
Eq.~(\ref{gluinodecay}).  In Subsection~\ref{secbr} we proposed to
investigate $\nfit$/$\nall$ where $\nfit$ is the number of the edge
event estimated by the fit and $\nall$ is the total $tb$ events
selected. It is related to the normalized branching ratio
$\Bredge/\Br(\tilde{g}\rightarrow bbX)$.  The normalized branching
ratio is sensitive to the branching ratios for $\stp$ and $\sbt$, and
uncertainty on fragmentation and $b$-tagging efficiency would be
canceled out in the ratio.

In Fig.~\ref{msugrafig3}(b) we show the branching ratio
Br$(\tilde{g}\rightarrow bbX)$ and the branching ratio of the gluino
to the edge events normalized by Br$(\tilde{g}\rightarrow bb X)$. They
are decreasing functions of $A_0$, since the $\stp$ and $\sbt$ is
heavier for larger $A_0$. While the behavior of
Br$(\tilde{g}\rightarrow bb X)$ is moderate, the normalized branching
ratio of the edge events is sensitive to $A_0$, since it depends on
the branching ratios of $\stp$ and $\sbt$. In Fig.~\ref{msugrafig3}(b)
the normalized branching ratio has two kinks, where some decay modes
for $\stp$ or $\sbt$ are open.  The first kink around $A_0\sim -
500$~GeV comes from a change of the branching ratio of $\stp$. The
dominant decay modes of $\stp$ are $b\chapm $ and $t\neu $, however,
the mode $t \neus $ is open around $A_0\sim -500$~GeV.  The decay
modes $\stp \to b\chaspm $ and $\sbt \to t\chaspm $ are open at the
second kink around $A_0\sim -150$~GeV.  The decay modes are not
included in the edge events, because the end point $M_{tb}({\rm
III})_{2}$ is significantly lower than $M_{tb}({\rm III})_{1}$.  The
additional tags for $\chaspm$ discussed in the previous section to
identify the decay works for the region. Since $\chaspm$ is
higgsino-like and the Yukawa coupling to $\sbt$ is enhanced by the
top-Yukawa coupling, the decay mode $\sbt \to t\chaspm $ becomes
dominant for $A_0>-150$~GeV, and then the normalized branching ratio
become quickly diluted.

\begin{figure}[ht]
\centerline{
\psfig{file=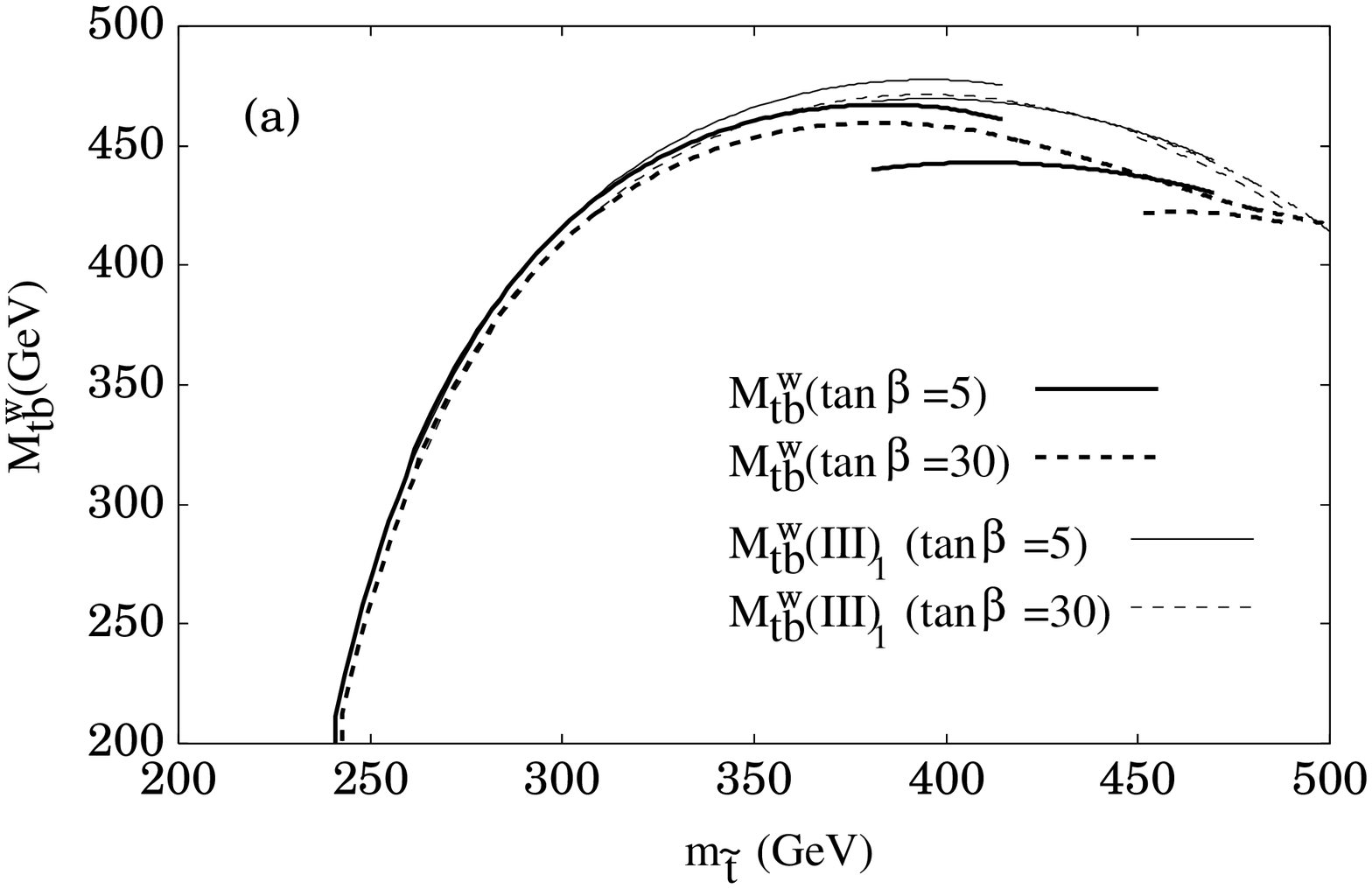,width=3in}}
\centerline{
\psfig{file=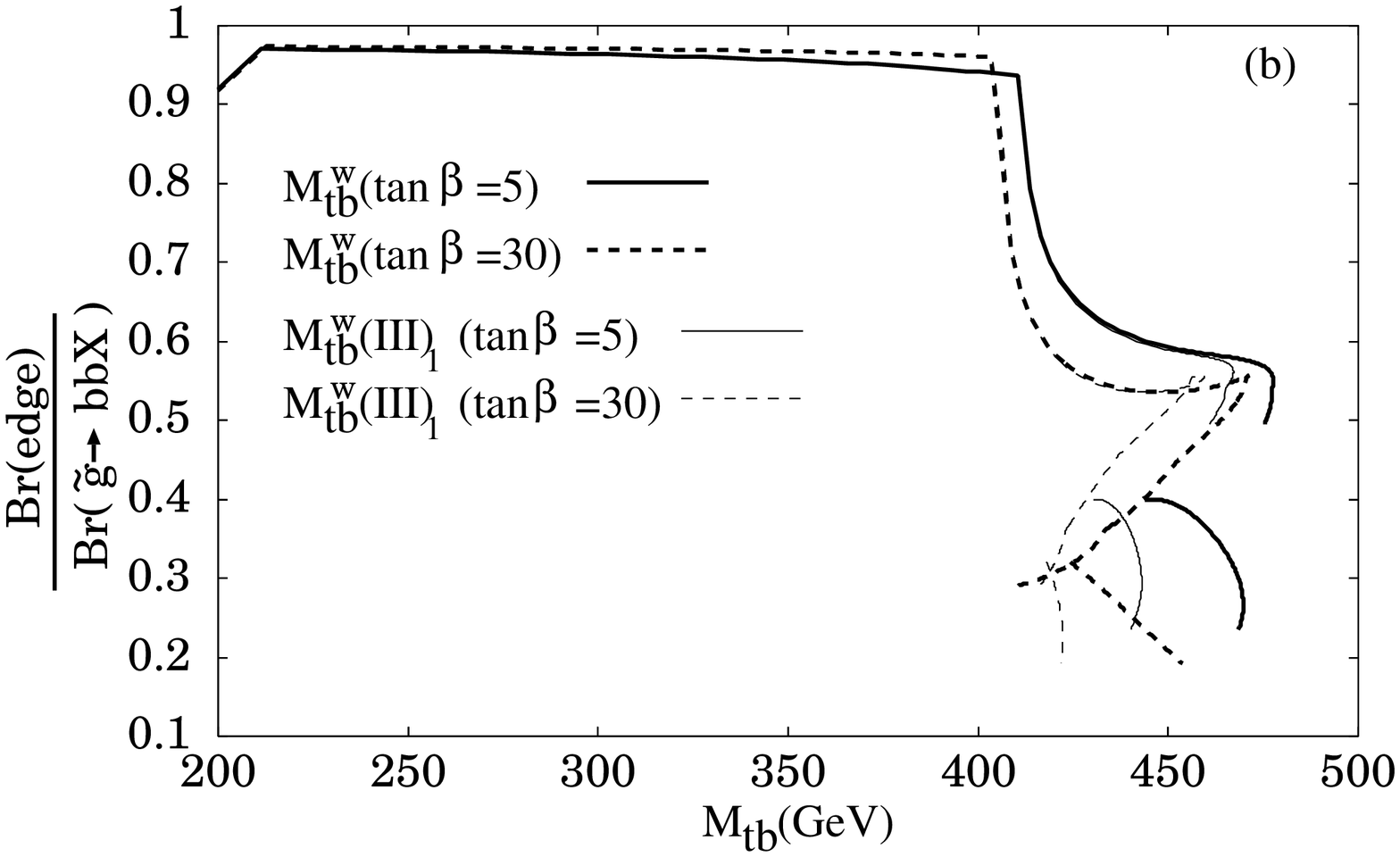,width=3in}
}
\caption{For fixed $m_{\sbt}=570$~GeV and $M_{\tilde{g}}=707$~GeV, 
(a) $M_{tb}^{\rm w}$ and $m_{\stp}$, and (b) $M_{tb}^{\rm w}$
$(M_{tb}{\rm (III)})$ and $\Bredge/\Br(\tilde{g}\rightarrow
bb X)$.  Here, $-2$~TeV~$<A_0<$~2~TeV, $\tan\beta=5$ and 30, and
$\mu>0$. $m_0$ is fixed by $m_{\sbt}$.  The neutralino LSP is
assumed.}
\label{msugrafig4}
\end{figure}

In the MSUGRA, the masses of ${\stp}$ and ${\sbt}$ are related with
each other in a broad parameter space.  As the determination of
$m_{\sbt}$ has been studied in Ref. \cite{TDR}, we now discuss $\mtbw$
and the normalized branching ratio with a fixed $m_{\sbt}$ value.  We
fix $m_{\sbt}=570$~GeV and $M_{\tilde{g}}=707$~GeV for $\tan\beta=5$
and 30, and show $M_{tb}^{\rm w}$ $(M_{tb}{\rm (III)_1})$ as a
function of $m_{\stp}$ in Fig.~\ref{msugrafig4}(a).  Our parameter
scan is restricted to the range $\vert A_0\vert < 2$~TeV.

We have two disconnected solutions corresponding to $-1400\
(-1850)$~GeV$<A_0<-280\ (470)$~GeV and $1350\ (1600)$~GeV
$<A_0<2000$~GeV for $\tan\beta=5\ (30)$, respectively.  The $A_0$
region between the two solutions is not allowed because of the
experimental constraint of $\tilde{\tau}$ mass or charged LSP.  For
$\tan\beta$=5 (30), $m_0$ is smaller than 460 (750)~GeV for $A_0<0$,
and 500 (300)~GeV for $A_0>0$.  Since $\sbt\sim \tilde{b}_L$ in the
MSUGRA, a larger $|A_0|$ corresponds to a larger $m_0$ for a fixed
$m_{\sbt}$, as expected from Eq.~(\ref{rg}). Nevertheless $\stp$ could
be much lighter than $\sbt$ for a large and negative value of $A_0$ as
can be seen in Fig.~\ref{msugrafig4}(a)
\footnote{ $\vert A_0 \vert < 3m_0$ for $A_0<0$, satisfying 
a condition for the consistent minimum for the potential.}. 
Note that $\mtbw$ is determined for a fixed 
$\tilde{t}_1$ mass  when $m_{\stp}$ is lighter than 370~GeV. 
If $m_{\sbt}$ is constrained elsewhere, $m_{\stp}$ is 
strongly restricted  by the $\mtbw$ measurement
under the MSUGRA assumption. 

In Fig.~\ref{msugrafig4}(b) we show the solution in a $\mtbw$ and
$\Bredge/\Br(\tilde{g}\rightarrow bbX)$ plane.  The normalized
branching ratios is almost 1 for $\mtbw<400$~GeV (or $\mstp<300$~GeV).
When the mass difference of $m_{\stp}$ and $m_{\sbt}$ is large, the
decay of $\sbt$ is dominated by $W\stp$. Furthermore, for
$m_{\stp}\lsim 300$~GeV, Br$(\stp\rightarrow b\chapm)$ is 100\%.
However since $m_{\tilde{q}}>m_{\tilde{g}}$ in this region, squark and
gluino production goes to the final states having four bottom quarks,
where we have seen the disagreement between the measurement
$\nedge/\nall$ and $\Bredge$/Br($\tilde{g}\rightarrow bbX$) in
Section~III~D.  In this case our study must be extended to events with
more than three tagged $b$-jets.

Note that the decay $\tilde{\chi}^0_2\rightarrow 
\tilde{l}$ is mostly closed in Fig.~\ref{msugrafig4}.
The decay is open only at the most right region of
Fig.~\ref{msugrafig4}(b), near the end of the lines of the
solutions. If this decay is open, the masses of $\tilde{q}, \tilde{l}$
and $\tilde{\chi}^0_1$ are model-independently determined by the
$m_{ll}$, $m_{llj}$ and $m_{lj}$ distributions.  When this decay mode
is kinematically forbidden, one needs to combine various distributions
for consistency check of the MSUGRA assumptions.  As is already shown
in Fig.~\ref{msugrafig2}, the decay $\tilde{g}\rightarrow
\tilde{b}/\tilde{t}$ is open up to $m_0\sim m_{\tilde{g}}$, providing
information of the third generation squarks in the wide region of the
parameter space.

\section{Measurement of Top Polarization}

\begin{figure}[ht]
\centerline{
\psfig{file=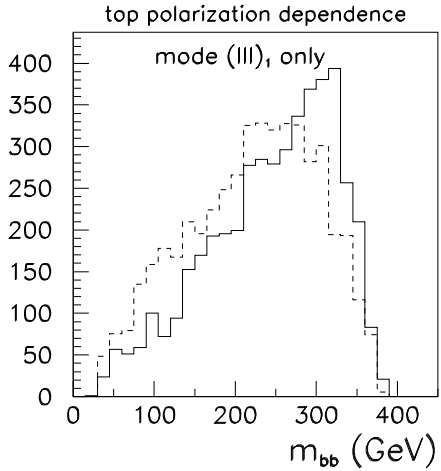,width=3in}}
\caption{
Distribution of $m_{bb}$ in the decay chain (III)$_1$.  The
(dashed) line is for $\stp=\tilde{t}_L(\tilde{t}_R)$, and
400~GeV$<m_{tb}<470$~GeV. 
We use the mass spectrum in the sample point A1 in Table~\ref{masscross}, 
and the normalization is arbitrary.
 }
\label{polfig1}
\end{figure}

\begin{figure}[ht]
\centerline{
\psfig{file=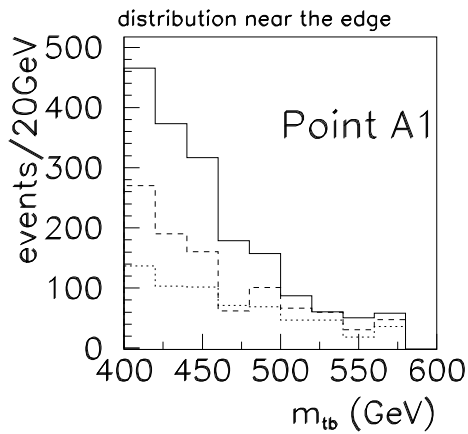,width=3in} 
}
\caption{
The solid line is for the total $m_{tb}$ distribution at the sample point
A1, the dashed line is for the $m_{tb}$ distribution excluding the mode
(III)$_1$, and the dotted line is for the $m_{tb}$ distribution excluding
the modes (III)$_1$, (III)$_{11}$ and (III)$_{21}$. }
\label{polfig2}
\end{figure} 

Similar to the tau-lepton decay, we may measure the polarization of
the top quark since it decays to $b W$ via the $(V-A)$ interaction. The
top quarks from the $\tilde{g}$, $\tilde{t}$, and $\tilde{b}$ decays are
polarized, and the polarization
depends on the mixing angles for stops, charginos,
and neutralinos.

The bottom quark angular distribution in the polarized top quark decay 
is the following,
\begin{equation}
\frac{1}{\Gamma_t}\frac{d\Gamma_t}{d\cos\theta}\propto
{\left(\frac{m_t}{m_W}\right)^2 \sin^2\frac{\theta}{2}
+2 \cos^2\frac{\theta}{2}},
\end{equation}
where $\theta$ is the angle between the direction of the bottom quark
and the direction of the top quark spin in the rest frame of the top
quark. The terms
proportional to $({m_t}/{m_W})^2$ come from the decay to the
longitudinal $W$ boson.  The bottom quark thus tends to go to the
opposite direction of the top quark spin.

In the decay mode (III)$_1$, the top quark from a gluino decay is
polarized to be left-handed (right-handed) if $\stp$ is left-handed
(right-handed).  The polarization is reflected on the invariant mass
distribution of the $bb$ system ($m_{bb}$), if the top quark is
relativistic enough in the gluino rest frame.  Here, one of the bottom
quarks comes from the top quark decay and the other comes from
$\stp\rightarrow b\tilde{\chi}^{\pm}$.  Especially, when the invariant
mass $m_{tb}$ is close to the end point $M_{tb}$ of the decay mode
(III)$_1$, the top and bottom quarks go to the opposite direction to
each other in the gluino rest frame.  Thus, the distribution of the
invariant mass $m_{bb}$ for events with $m_{tb}$ close to $M_{tb}({\rm
III})_1$ is harder (softer) for left-handed (right-handed) top quarks.

In Fig.~\ref{polfig1} we show the $m_{bb}$ distribution from the decay
chain (III)$_1$. In this simulation, we use the HERWIG generator,
since it respects helicities for each particles in the processes.  We
generated a large number of events which go through the decay
(III)$_1$.  We use the mass spectrum at the reference point A1 in
Table~\ref{masscross}.  We use events with 400~GeV$<m_{tb}<470$~GeV to
make the distribution, and the solid (dotted) line is for the
left-handed (right-handed) stop.  The statistical significance in the
difference between the left-handed and right-handed stops is about
$3\sigma$ for ${\cal O}(100)$ events.

In the above simulation, we neglect contribution of other decay chains
such as (III)$_{11}$ and (IV)$_{11}$.  The other decay chains may
contribute to the $m_{bb}$ distribution even if we impose that
$m_{tb}$ is near the end point. The $m_{bb}$ distributions in the
modes (III)$_{11}$ and (IV)$_{11}$ do not depend on the polarization
for the top quark, since the top quark distribution in the scalar
boson decay is spherical. In Fig.~\ref{polfig2} we show the $m_{tb}$
distribution around the edge region at the reference point A1.  The
solid line is for the total distribution, the dashed line is for the
distribution excluding the mode (III)$_1$, and the dotted line is for
the distribution excluding the modes (III)$_1$, (III)$_{11}$, and
(III)$_{21}$.  About a half of the events near the end point come from
the signal mode (III)$_{1}$, and the ratio between the signal events
and the rest depends on the MSSM parameters.

\section{Conclusions}

In this paper we study
cascade decays $\tilde{g}\rightarrow (t\tilde{t}_1\ {\rm or } \ b\tilde{b}_i)
\rightarrow 
tb\tilde{\chi}^{\pm}_i$ at the LHC by reconstructing $tb$ final state
where the top quark decays hadronically.  The $m_{tb}$ distribution of
the cascade decay has an edge structure.  The measurement of the end
point and the edge height of the $m_{tb}$ distribution constrains a
combination of the masses of $\tilde{g}, \tilde{b}, \tilde{t}$ and
$\tilde{\chi}^{\pm}$ and the decay branching ratios of the particles
involved in the decays.

Through a detailed simulation study, we show in this paper the
measurement of the end point and edge height on a continuum background
is indeed possible. Namely, the end point of the cascade decay
calculated in parton level agrees with the reconstructed edge
position, and the ratio of $\nedge$ (the number of reconstructed edge
events) and $\nall$ (the number of total reconstructed $tb$ events) is
understood well by the ratio $\Bredge/\Br(\tilde{g}\rightarrow bbX)$.

The end point and branching ratios depend on the mass and left-right
mixing of the $\tilde{t}$ and $\tilde{b}$, as well as the chargino and
neutralino masses and mixings in the MSSM. In the MSUGRA, these
sparticle spectrum is expressed by a few parameters at the GUT
scale. The decay mode $\tilde{g}\rightarrow (t\tilde{t}_1\ {\rm or} \
b\tilde{b}_i)\rightarrow tb\tilde{\chi}^{\pm}_1$ is open for a wide
region of the parameter space where $m_0<m_{\tilde{g}}$.  The $m_{tb}$
distribution is sensitive to the $A_0$ parameter, the trilinear
coupling at the GUT scale. The distribution is most sensitive to $A_0$
when $A_0 \cdot M_0<0$.

The stop and sbottom  could 
decay both into the heavier and lighter charginos and neutralinos
unlike the first and second generation squarks. This is  
because the third generation squarks have the large top (bottom)-Yukawa
coupling to the higgsinos. A strategy to  search for   
such  decays  specific for the stop and sbottom  
is described in Section~IV. The Yukawa coupling is also related to the  
$\tilde{t}_L$-$\tilde{t}_R$ mixing. The polarization of the top 
quark arising from the gluino decay $\tilde{g}\rightarrow t\tilde{t}$
depends on the stop left-right mixing. The dependence of the 
$m_{bb}$ distribution on the top polarization in the tagged $tb$ 
sample is discussed in Section~VI. 
 
To understand the event distribution better, one needs to 
know the nature of the  quark and gluon  fragmentations into jets. 
The reconstruction efficiencies are significantly different between 
the two standard SUSY generators HERWIG and PYTHIA, which
adopt  different 
models for the fragmentation. We point out that 
the smearing of the jet-pair invariant mass 
originating from a $W$ decay affects the 
reconstruction efficiency.

The interplay between the LHC and the future LC would be useful to
reduce the systematics coming from the uncertainties of sparticle
masses and decay branching ratios. They would be reduced dramatically
if some of the charginos and neutralinos are accessible at the LC.
Model independent and precise determination of the stop and sbottom
masses may be possible in such cases and will be discussed elsewhere.

\section*{Acknowledgment}

We thank the ATLAS collaboration members for useful discussion. We
have made use of the physics analysis framework and tools which
are the result of collaboration-wide efforts.
We especially  thank Dr. Kanzaki and Mr. Toya. We acknowledge ICEPP,
Univ. of Tokyo, for providing us computing resources.
This work is
supported in part by the Grant-in-Aid for Science Research, Ministry
of Education, Science and Culture, Japan 
(No.13135297 and  No.14046225 for JH,
No.11207101 for KK, and 
No.14540260 and 14046210 for MMN).

\section*{Appendix:  Reliability of Sideband Subtraction}

In this paper, we generate events both by the PYTHIA and HERWIG
generators.  The HERWIG generator uses a parton-shower approach for
initial- and final- state QCD radiations including the color coherence
effects and the azimuthal correlation both within and between jets
\cite{HERWIG}.  The full available phase space for the parton shower
is restricted to an angular order region, namely, the angle between
the two emitted partons is smaller than that of previous branches.  On
the other hand, the PYTHIA generator adopts the string model
\cite{PYTHIA}.  The two generators predict different $tb$
reconstruction efficiencies, which may be considered as the
uncertainty in fragmentation.

Definition of jets also affects the number of 
reconstructed $tb$ events and the reconstruction efficiency. 
We try two algorithms, a cone-based algorithm and 
a $K_T$ algorithm (Montreal version), available in 
the JET Finder Library~\cite{jetfinder}, 
which is interfaced to the ATLFAST packages. 

The cone-based algorithm merges the calorimeter cells around a high
$E_T$ cell in a fixed cone size $R$.  On the other hand, in the $K_T$
algorithm, a cell called a protojet $i$ grows until there are no more
protojets $j$ which satisfy $(\Delta \eta)^2+ (\Delta \phi)^2<(
E^2_{Ti}/{\rm min}(E^2_{Ti}, E^2_{Tj})) R^2$, therefore the cone size
depends on $E_T$'s, and the shape of the jet depends on the
distribution of the particles in the jet.  The $K_T$ algorithm is
based on QCD, and it is considered to be advantageous to merge soft
jets from an initial parton, although experimentally challenging.

The fit results with different generators and jet finding algorithms
at the point A1 is summarized in Table~\ref{fitjet}.  The end point
$\mtbfit$ becomes closer to $\mtbw$ as $R$ increases. This is expected
because more soft jets are merged to leading jets as $R$ increases.
The edge height $h$ is larger for HERWIG with the $K_T$ algorithm.
The difference of the height for $R\leq 0.5$ is more than 25\% in
Table~IX.  The height significantly decreases for $R=0.6$. This is
because we have to reconstruct a jet pair with relatively small
invariant mass $\sim$ 80~GeV.

\begin{table}[t]
\begin{tabular}{|cr|cc|}
\hline
gen&	jet&	$\mtbfit$[GeV]&  $h$/(10~GeV)\cr
\hline
PY&	cone 0.4&	455.2 $\pm$ 8.2 &271.4 $\pm$ 22.7	\cr
&	0.5&	436.1 $\pm$ 7.7 &272.9 $\pm$ 33.4	\cr
&	0.6&	461.1 $\pm$ 10.4 &217.8 $\pm$ 22.8 \cr
&	$K_T$ 0.4&	442.0 $\pm$ 4.7&321.3 $\pm$ 22.3\cr
&	0.5&	452.2 $\pm$ 6.0&	305.3 $\pm$ 21.3\cr
&	0.6&	459.1 $\pm$ 6.1&	241.6 $\pm$ 31.6\cr
\hline
HW&	cone 0.4& 434.5 $\pm$ 5.8& 354.8 $\pm$ 23.3\cr
&	   0.5&	460.2 $\pm$ 4.9& 349.2 $\pm$ 22.8\cr
&	0.6&	440.9 $\pm$ 7.1& 305.3 $\pm$ 33.7\cr
& $K_T$ 0.4&	434.9 $\pm$ 4.3& 406.5 $\pm$ 22.1\cr
&	0.5&	460.0 $\pm$ 5.5& 379.6 $\pm$ 33.5\cr
&	0.6&	468.4 $\pm$ 5.8& 314.3 $\pm$ 20.7\cr
\hline
\end{tabular}
\caption{Fit results at the point A1. 
We use the cone-based algorithm or the $K_T$
algorithm with cone sizes of $R=0.4, 0.5, 0.6$.  
The fit is based on $3\times 10^6$ events. The 
$\mtbw$ is 459~GeV. }
\label{fitjet}
\end{table}

\begin{table}
\begin{tabular}{|cr||rr|rr|}
\hline
 &&HW&&PY&\cr
 $K_T$/cone & R& sig& after sub & sig& after sub 
\cr
\hline
{\bf mode (III)$_1$}& &&&&\cr
cone &0.4 & 10180  &  2949  &  9695  &  2462 \cr
cone &0.5 &  9659  &  2974 &  9213  &  2447 \cr
$K_T$ &0.4 &  9405  &  3105 &  9226  &  2813\cr
$K_T$ &0.5 &  9118  &  3239 &  8607  &  2808\cr
\hline
{\bf mode (I)$_2$}&&&&&\cr
cone &0.4 & 2779 & 574& 2630& 500\cr
$K_T$ &0.5 & 2086& 448& 2058& 433\cr
\hline
\end{tabular}
\caption{Numbers of events before and after the sideband 
subtraction for the events with a gluino that decays through 
mode (III)$_1$ 
$\tilde{g}\rightarrow t\tilde{t}_1\rightarrow tb\tilde{\chi}^{\pm}_1$
and   mode (I)$_2$ 
$\tilde{g}\rightarrow b\tilde{b}_1\rightarrow bb\tilde{\chi}^0_2$.}
\label{mode}
\end{table}

\begin{figure}
\centerline{\psfig{file=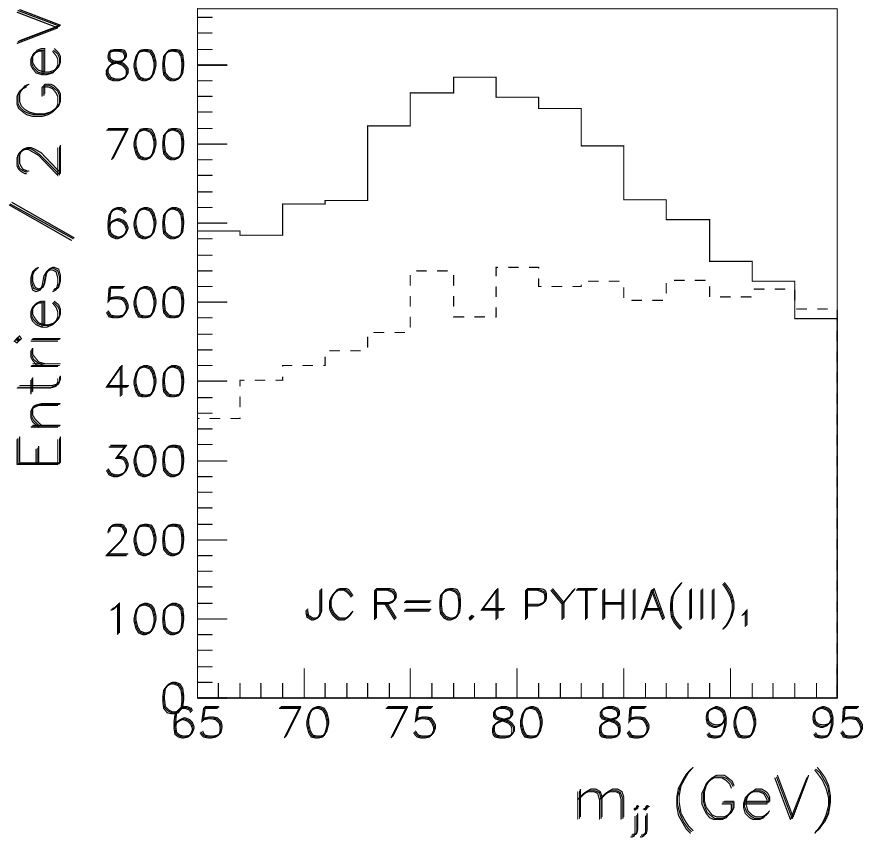,width=2in} 
\hskip -1cm\psfig{file=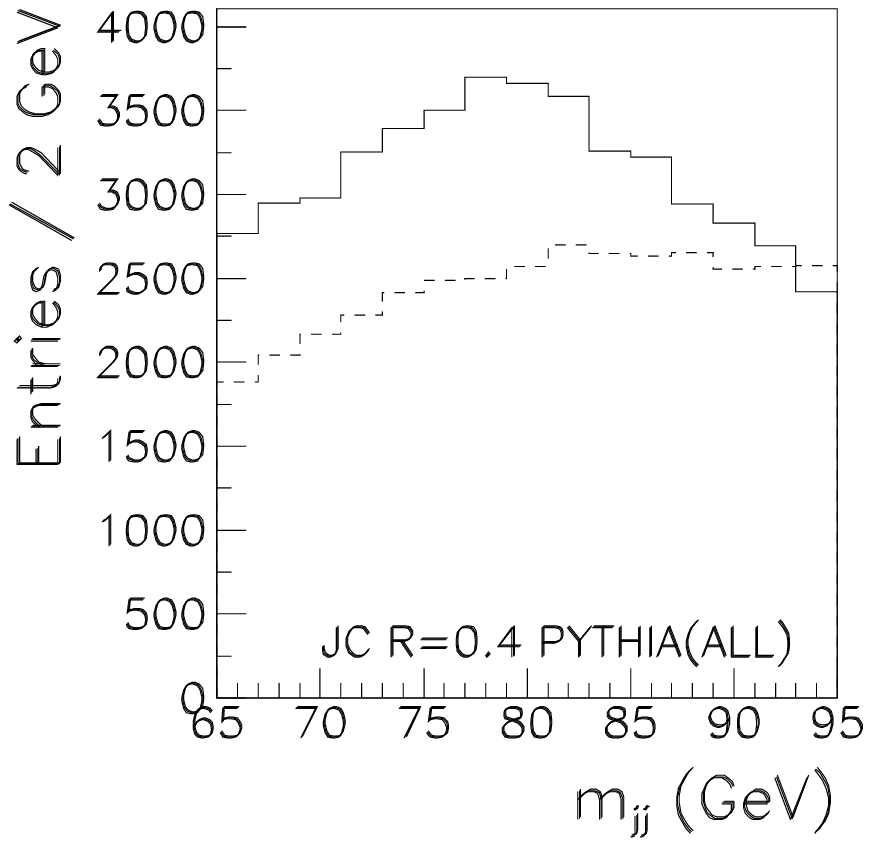,width=2in}} 
\centerline{\psfig{file=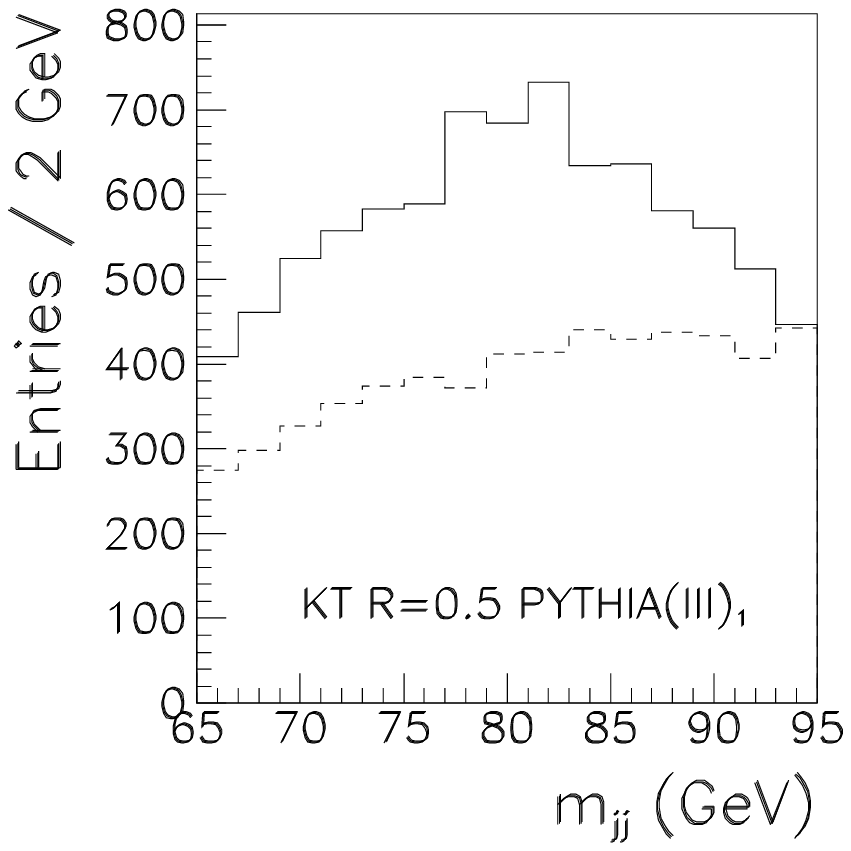,width=2in} 
\hskip -1cm \psfig{file=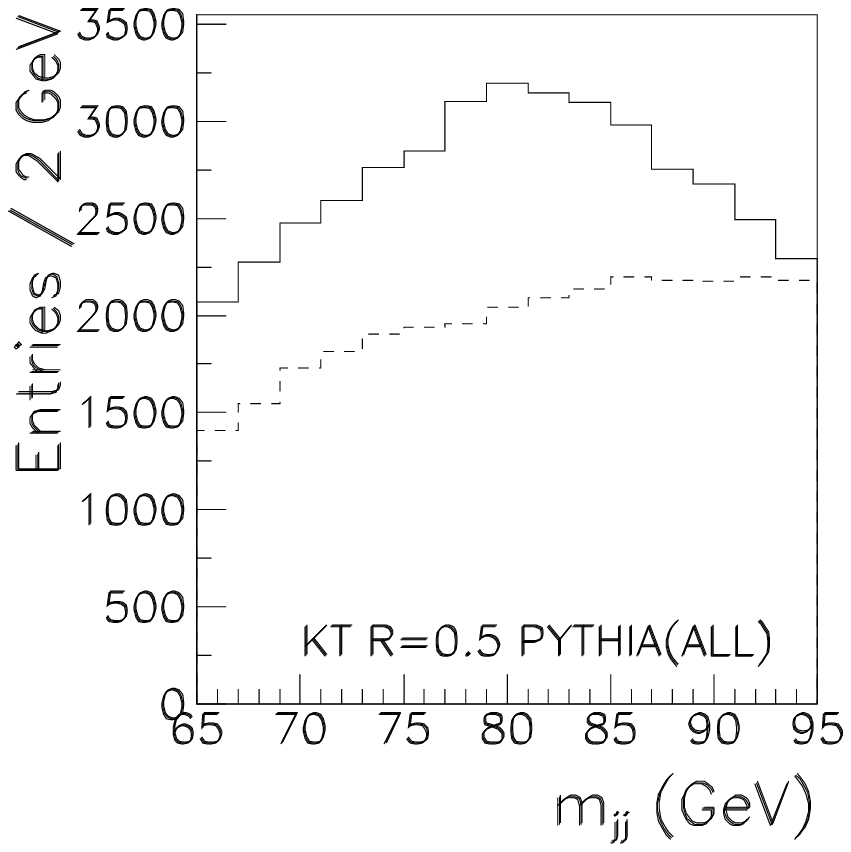,width=2in}}
\centerline{\psfig{file=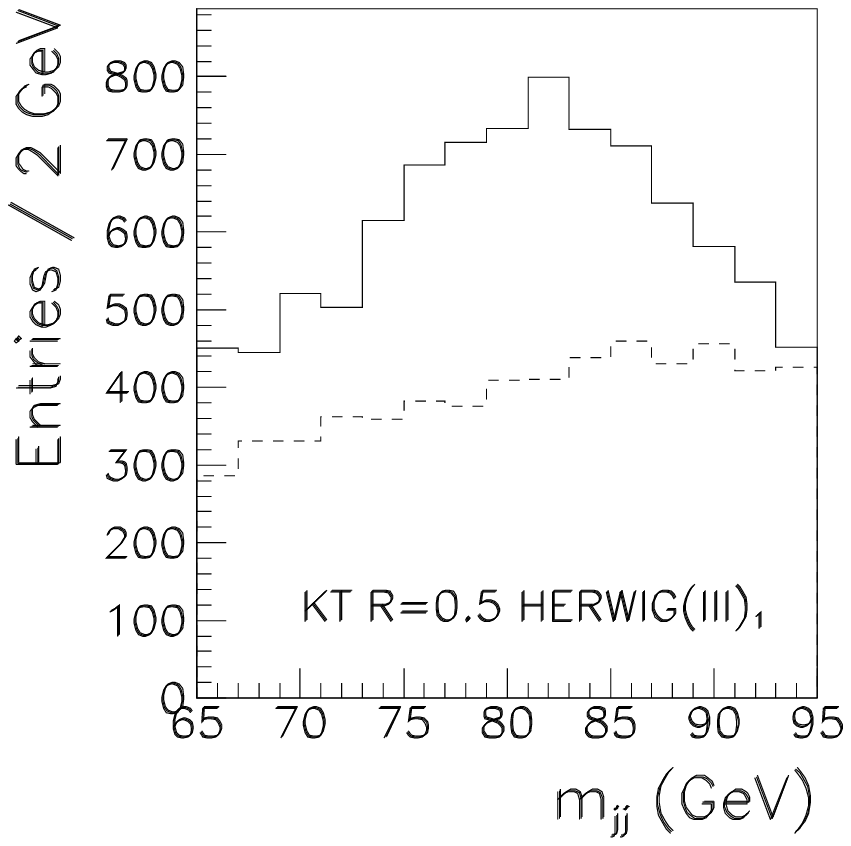,width=2in}
\hskip -1cm \psfig{file=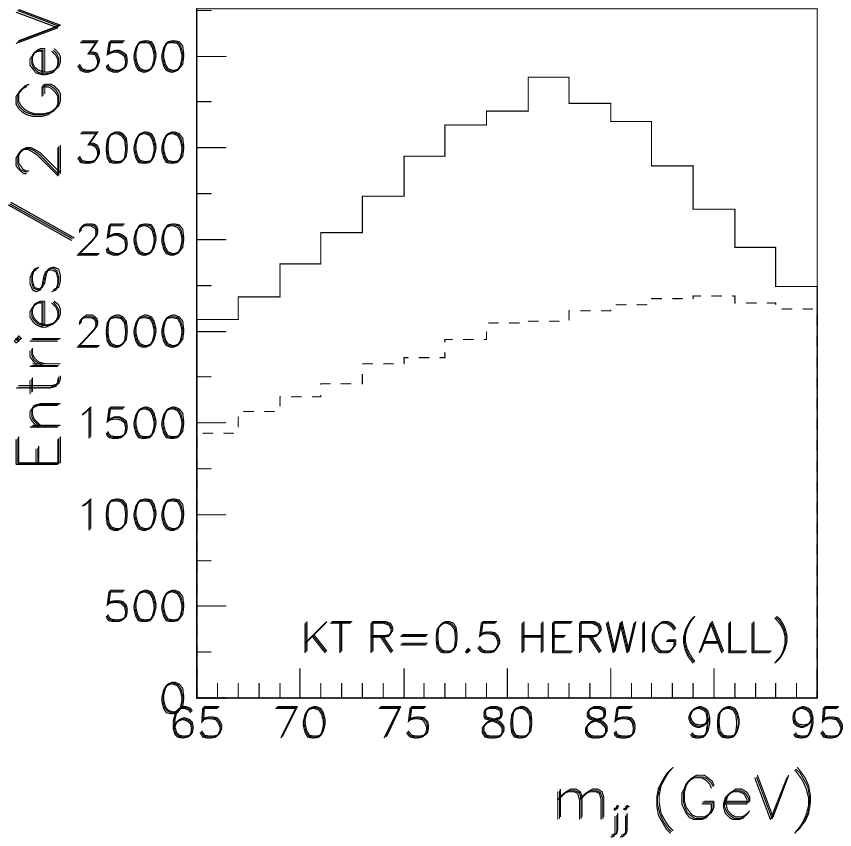,width=2in}}
\caption{ Distributions of $m_{jj}$ consistent to 
the $W$ boson mass under the cuts described in Section~II.  
The MC data is 
generated for the point A1. In the left, we plot the events that contain 
the decay chain  
$\tilde{g}\rightarrow t\tilde{t}_1\rightarrow tb\tilde{\chi}^{\pm}_1$
only, while we plots all events in the right. 
 Dotted lines show estimated background. 
Top figures: PYTHIA with the cone-based algorithm ($R=0.4$). 
Middle: PYTHIA with the $K_T$ algorithm ($R=0.5$). 
Bottom: HERWIG with the $K_T$ algorithm ($R=0.5$).}
\label{modemjj}
\end{figure}

To see the origin of the difference more closely, we compare the
$m_{jj}$ distributions consistent to the top interpretation in
Fig.~\ref{modemjj}, where, the solid histograms are the invariant mass
distributions for jet pairs which satisfy $\vert
m_{jj}-m_W\vert<15$~GeV and $\vert m_{bjj}-m_t\vert<30$~GeV.  Dashed
histograms show the distribution of accidental jet pairs in the $W$
mass region, that is, the ``fake $W$'' background, estimated by the
sideband method.  The left plots are the distributions of the events
arising from the decay chain (III)$_1$, while the right plots are the
$m_{tb}$ distribution of all selected $tb$ events.  The distribution
at the bottom-left (HERWIG and the $K_T$ algorithm for $R=0.5$ ) are
more concentrated around $m_{jj}\sim 80$~GeV compared to the others
(PYTHIA and the cone-based or the $K_T$ algorithm), corresponding to
better reconstruction efficiencies.

Incompleteness of the sideband subtraction is seen in the same figure.
The sideband subtraction uses the jet pairs with invariant mass
35~GeV$<m_{jj}<65$~GeV and 95~GeV$<m_{jj}<$~125~GeV to estimate the
``fake W'' background. On the other hand, as the jets are defined to
have $p^{jet}_T>$10~GeV, a jet pair near $m_{jj}\sim 35$~GeV is
suppressed.  We see an underestimate of background events in the range
65~GeV$<m_{jj}<$ 70~GeV, where the dotted line and solid line differ
significantly.  We find that events with $m_{jj}<70$~GeV do not
contribute to the edge structure of the $m_{tb}$ distribution.  The
events in this region might have to be removed from the candidate
events.

The incompleteness of the sideband subtraction can also be seen by
comparing the reconstruction efficiency of the events from specific
decay chains.  Here we take two decay chains for comparison;
\begin{eqnarray}
{\rm (III)_1} && \glu \rightarrow t\stp  \rightarrow tb\chapm,\cr
{\rm (I)_2} && \glu \rightarrow b\sbt  \rightarrow bb\neus. 
\end{eqnarray}
The  decay chain (III)$_1$ is relevant to the $tb$ events.  
On the other hand, as  the decay chain 
(I)$_2$ has no $W$, 
the sideband subtraction should remove the events arising from 
the decay chain completely, unless there are accidental weak bosons, 
or the understimation of the misreconstructed events that 
has just been discussed above. 

The branching ratio 
$\Br(\tilde{\chi}^0_2\rightarrow Z^0\tilde{\chi}^0_1)$ is small 
(1.5\%) at the point A1.
The  decay branching ratios at the point A1 are
\begin{eqnarray}
\Br({\rm III})_1&=& \Br(\glu\rightarrow t\stp ) 
\Br(\stp \rightarrow b\chapm)\times 0.7
\cr
&=&0.078 \nonumber \\
\Br({\rm I})_2 &=& \Br(\glu\rightarrow b\sbt) \Br(\sbt\rightarrow b\neus)
\cr
&=& 0.025, 
\end{eqnarray}
where the factor 0.7 is the hadronic branching ratio of the $W$ boson.

The numbers of reconstructed $tb$ events before and after the sideband
subtraction are listed in Table~\ref{mode}.  For the $K_T$ algorithm
with $R=0.5$, the ratio of the numbers N(I)$_2$/N(III)$_1$ are 0.229
and 0.15 before and after the sideband subtraction, respectively.
Comparing these ratios to that of the branching ratios
Br(I)$_2$/Br(III)$_1$=0.32, we see that the requirement $| m_{bjj} -
m_t | < 30$~GeV reduces the contribution of (I)$_2$ to about 70\%
relative to (III)$_1$ and it is further reduced to less than 50\%
after the sideband subtraction. This ratio may be improved further by
cutting the events with $m_{jj}<70$~GeV or by using more sophisticated
background estimation.

\end{document}